\documentstyle[sprocl,epsfig,psfig]{article}

\documentstyle[sprocl,epsfig,psfig]{article}

\def\be{\begin{equation}}
\def\ee{\end{equation}}
\def\bea{\begin{eqnarray}}
\def\eea{\end{eqnarray}}

\newcommand{\newc}{\newcommand} 
\newc{\lra}{\leftrightarrow} 
\newc{\beq}{\begin{equation}} 
\newc{\eeq}{\end{equation}} 
\newc{\barr}{\begin{eqnarray}} 
\newc{\earr}{\end{eqnarray}} 

\begin{document} 
\title{THE MODULATION EFFECT FOR SUPERSYMMETRIC DARK MATTER DETECTION
WITH ASYMMETRIC VELOCITY DISPERSION. }

\author{J. D. VERGADOS} 

\address{Theoretical Physics Section, University of Ioannina, GR-45110, 
Greece\\E-mail:Vergados@cc.uoi.gr} 

\maketitle\abstracts{  The detection of the theoretically expected dark matter
is central to particle physics cosmology. Current fashionable supersymmetric
models provide a natural dark matter candidate which is the lightest
supersymmetric particle (LSP). Such models conmbined with fairly well 
understood physics like
the quark substructure of the nucleon and the nuclear form factor and/or the
spin response function of the nucleus, permit the evaluation of
the event rate for LSP-nucleus elastic scattering. The thus obtained event rates
are, however, very low or even undetectable.
 So it is imperative to exploit the modulation effect, i.e. the dependence of
the event irate on  the earth's annual motion.  In this paper we study
such a modulation effect both in non directional and directional experiments.
We calculate both the differential and the total rates using symmetric as well
as asymmetric velocity distributions. We find that in the symmetric case 
the modulation amplitude is small, less than 0.07. There exist, however, 
regions of 
the phase space and experimental conditions such that the effect  can become
larger. The inclusion of asymmetry, with a realistic enhanced velocity 
dispersion  in the galactocentric direction, yields the bonus of an enhanced
modulation effect, with an amplitude which for certain parameters 
can become as large as 0.46}
\section{Introduction}
It is known that that dark matter is needed to close the Universe
\cite {KTP}, \cite{Jungm}.
It is also known that one needs two kinds of dark matter. One composed of 
particles which were relativistic at the time of structure
formation. These  constitute the hot
dark matter component (HDM). The other is made up of particles which were 
non-relativistic at the time of freeze out. This is  the cold dark 
matter component (CDM). The COBE data ~\cite{COBE} suggest that CDM
is at least $60\%$ ~\cite {GAW}. On the
other hand recent data from the Supernova Cosmology Project suggest ~\cite {SPF}
,~\cite {SCP} that there is no need for HDM and the situation can be adequately
described by $\Omega<1$, e.g. $\Omega _{CDM}= 0.3$ and $\Omega _{\Lambda}= 0.6$.
In a more recent analysis Turner \cite {Turner} gives $\Omega_{m}=0.4$.

Since the non exotic component cannot exceed $40\%$ of the CDM 
~\cite{Jungm}$^,$~\cite {Benne}, there is room for the exotic WIMP's 
(Interacting Massive Particles).  Recently the DAMA experiment ~\cite {BERNA2} 
has claimed the observation of one signal in direct detection of a WIMP, which
with better statistics has subsequently been interpreted as a modulation signal
~\cite{BERNA1}.

In the currently favored supersymmetric
extensions of the standard model the most natural WIMP candidate is the LSP,
i.e. the lightest supersymmetric particle, whose nature can be described in 
most supersymmetric (SUSY) models to be a Majorana fermion, a linear 
combination of the neutral components of the gauginos and Higgsinos
~\cite{JDV}$^-$\cite{Hab-Ka}. 

 Since this particle is expected to be very massive, $m_{\chi} \geq 30 GeV$, and
extremely non relativistic with average kinetic energy $T \leq 100 KeV$,
it can be directly detected ~\cite{JDV}$^-$\cite{KVprd} only via the recoiling
of a nucleus (A,Z) in the elastic scattering process:
\begin{equation}
\chi \, +\, (A,Z) \, \to \, \chi \,  + \, (A,Z)^* 
\end{equation}
($\chi$ denotes the LSP). In order to compute the event rate one proceeds
with the following steps:

1) Write down the effective Lagrangian at the elementary particle 
(quark) level obtained in the framework of supersymmetry as described 
in Refs.~\cite{Jungm}, Bottino {\it et al.} \cite{ref2} and \cite{Hab-Ka}.

2) Go from the quark to the nucleon level using an appropriate quark 
model for the nucleon. Special attention in this step is paid to the 
scalar couplings, which dominate the coherent part of the cross section
and the isoscalar axial current, which,
as we will see, strongly depend on the assumed quark model 
~\cite{Dree,Adler,Chen}

3) Compute the relevant nuclear matrix elements~\cite{Ress}$^-$\cite{Nikol}
using as reliable as possible many body nuclear wave functions hoping
that, by putting as accurate nuclear physics input as possible, 
one will be able to constrain the SUSY parameters as much as possible.

4) Calculate the modulation of the cross sections due to the earth's
revolution around the sun by a folding procedure
assuming some distribution~\cite{Jungm,Druk} of velocities for LSP.

 The purpose of our present review is to focus on the last point of our above 
list along the lines suggested by our recent letter \cite{JDV99}, expanding 
our previous results and giving some of the missing calculational details.
For the reader's convenience, however, we will give a brief description on the 
basic
ingredients on how to calculate LSP-nucleus scattering cross section, without
elaborating on how one gets the needed parameters from supersymmetry. For
the calculation of these parameters from  
representative input in the restricted SUSY parameter space, we refer the
reader to the literature, e.g. Bottino {\it et al.} ~\cite{ref2}, 
Kane {\it et al.} , Castano {\it et al.} and Arnowitt {\it et al.} \cite {ref3}.
Then we will specialize our study in the case of the nucleus $^{127}I$ which
is one of the most popular targets~\cite{Smith}$^-$\cite{BERNA2}. To
this end we will include the effect of the nuclear form factors.
We will consider both a symmetric Maxwell-Boltzmann distribution
\cite {Jungm} as well as asymmetric distributions like the one suggested by
Drukier \cite {Druk}. We will examine the effect modulation in the directional 
as well as the non directional detection, both in the differential as well
as the total event rates. We will present our results a function of the
LSP mass, $m_{\chi}$, for various detector energy thresholds, in a way
which can be easily understood by the experimentalists.

\section{The Basic Ingredients for LSP Nucleus Scattering}
 Because of lack of space we are not going 
to elaborate here further on the construction of the effective Lagrangian
derived from supersymmetry, but refer the reader to the literature \cite
{JDV,KVprd,ref1,ref2,KVdubna}. 
The effective Lagrangian can be obtained in first
order via Higgs exchange, s-quark exchange and Z-exchange. We will use
a formalism which is familiar from the theory of weak interactions, i.e. 
\beq
{\it L}_{eff} = - \frac {G_F}{\sqrt 2} \{({\bar \chi}_1 \gamma^{\lambda}
\gamma_5 \chi_1) J_{\lambda} + ({\bar \chi}_1 
 \chi_1) J\}
 \label{eq:eg 41}
\eeq
where
\beq
  J_{\lambda} =  {\bar N} \gamma_{\lambda} (f^0_V +f^1_V \tau_3
+ f^0_A\gamma_5 + f^1_A\gamma_5 \tau_3)N
 \label{eq:eg.42}
\eeq
and
\beq
J = {\bar N} (f^0_s +f^1_s \tau_3) N
 \label{eq:eg.45}
\eeq

We have neglected the uninteresting pseudoscalar and tensor
currents. Note that, due to the Majorana nature of the LSP, 
${\bar \chi_1} \gamma^{\lambda} \chi_1 =0$ (identically).
The parameters $f^0_V, f^1_V, f^0_A, f^1_A,f^0_S, f^1_S$  depend
on the SUSY model employed. In SUSY models derived from minimal SUGRA
the allowed parameter space is characterized at the GUT scale by five 
parameters,
two universal mass parameters, one for the scalars, $m_0$, and one for the
fermions, $m_{1/2}$, as well as the parameters 
$tan\beta $, one of $ A_0 $ and  $ m^{pole}_t $  and the
sign of $\mu $ \cite{ref3}. Deviations from universality at the GUT scale
have also been considered and found useful \cite{ref4}. We will not elaborate
further on this point since the above parameters involving universal masses
have already been computed in some models \cite{JDV,KVdubna} and effects
resulting from deviations from universality will be published elsewhere
\cite{WKV} (see also Arnowitt {\it et al} in Ref. \cite{ref4} and 
Bottino  {\it et al} in Ref. \cite{ref2}). For some choices in the allowed
parameter space the obtained couplings can be found in a previous paper 
\cite{KVdubna}.

The invariant amplitude in the case of non-relativistic LSP 
can be cast ~\cite{JDV} in the form
\barr
|{\cal M}|^2 &=& \frac{E_f E_i -m^2_x +{\bf p}_i\cdot {\bf p}_f } {m^2_x} \,
|J_0|^2 +  |{\bf J}|^2 +  |J|^2 
 \nonumber \\ & \simeq & \beta ^2 |J_0|^2 + |{\bf J}|^2 + |J|^2 
\label{2.1}
 \earr
where $m_x$ is the LSP mass, $|J_0|$ and $|{\bf J }|$ indicate the matrix 
elements of the time and space components of the current $J_\lambda$ 
of Eq. (\ref{eq:eg.42}), respectively, and $J$ represents the matrix 
element of the 
scalar current J of Eq. (\ref{eq:eg.45}). Notice that $|J_0|^2$ is multiplied
by $\beta^2$ (the suppression due to the Majorana nature of LSP mentioned
above). It is straightforward to show that 
\beq
 |J_0|^2 = A^2 |F({\bf q}^2)|^2 \,\left(f^0_V -f^1_V \frac{A-2 Z}{A}
 \right)^2
\label{2.2}
\eeq
\beq
 J^2 = A^2 |F({\bf q}^2)|^2 \,\left(f^0_S -f^1_S \frac{A-2 Z}{A}
 \right)^2
\label{2.3}
\eeq
\beq
|{\bf J}|^2 = \frac{1}{2J_i+1} |\langle J_i ||\, [ f^0_A {\bf
\Omega}_0({\bf q})\,+\,f^1_A {\bf \Omega}_1({\bf q}) ] \,||J_i\rangle |^2 
\label{2.4}
\eeq
with $F({\bf q}^2)$ the nuclear form factor and
\beq 
{\bf \Omega}_0({\bf q})  = \sum_{j=1}^A {\bf \sigma}(j) e^{-i{\bf q} \cdot
{\bf x}_j }, \qquad
{\bf \Omega}_1({\bf q})  = \sum_{j=1}^A {\bf \sigma} (j) {\bf \tau}_3 (j)
 e^{-i{\bf q} \cdot {\bf x}_j }
\label{2.5}  
\eeq
where ${\bf \sigma} (j)$, ${\bf \tau}_3 (j)$, ${\bf x}_j$ are the spin, third
component of isospin ($\tau_3 |p\rangle = |p\rangle$) and coordinate of
the j-th nucleon and $\bf q$ is the momentum transferred to the nucleus.

The differential cross section in the laboratory frame takes the form~\cite{JDV}
\barr
\frac{d\sigma}{d \Omega} &=& \frac{\sigma_0}{\pi} (\frac{\mu_r}{m_N})^2
\xi  \{\beta^2 |J_0|^2  [1 - \frac{2\eta+1}{(1+\eta)^2}
\xi^2 ] + |{\bf J}|^2 + |J|^2 \} 
\label{2.6}
 \earr
where $m_N$ is the proton mass, $\eta = m_x/m_N A$, $ $
$\xi = {\bf {\hat p}}_i \cdot {\bf {\hat q}} \ge 0$ (forward scattering) and  
\beq
\sigma_0 = \frac{1}{2\pi} (G_F m_N)^2 \simeq 0.77 \times 10^{-38}cm^2 
\label{2.7} 
\eeq
The reduced mass $\mu_r$ is given by
\beq
\mu _r = \frac{m_{\chi }}{1 + \eta}
\label{2.8} 
\eeq
For the evaluation of the differential rate, which is the main subject of the
present work, it will be more convenient to use
the variables $ (\upsilon ,u) $ instead of
the variables $ (\upsilon ,\xi) $. Thus 
integrating the differential cross section, Eq. (\ref {2.6}), with respect to 
the azimuthal angle we obtain
\beq
d\sigma (u,\upsilon) = \frac{du}{2 (\mu _r b\upsilon )^2} [(\bar{\Sigma} _{S} 
                       +\bar{\Sigma} _{V}~ \frac{\upsilon^2}{c^2}~F^2(u))
                       +\bar{\Sigma} _{spin} F_{11}]
\label{2.9}
\eeq
with 
\beq 
\bar{\Sigma} _{S} = \sigma_0 (\frac{\mu_r}{m_N})^2  \,
 \{ A^2 \, [ (f^0_S - f^1_S \frac{A-2 Z}{A})^2 \, ] 
\label{2.10}
\eeq 
\beq 
\bar{\Sigma} _{spin} = \sigma_0 (\frac{\mu_r}{m_N})^2  \,
[f^0_A \Omega_0(0))^2 \frac{F_{00}(u)}{F_{11}(u)} +
 2f^0_A f^1_A \Omega_0(0) \Omega_1(0)
\frac{F_{01}(u)}{F_{11}(u)}+ (f^1_A \Omega_1(0))^2  \, ] 
\label{2.11a}
\eeq 
\beq 
\bar{\Sigma} _{V}  =  \sigma_0 (\frac{\mu_r}{m_N})^2  \,
      A^2 \, (f^0_V - f^1_V \frac{A-2 Z}{A})^2 [ 1  
          -\frac{1}{(2 \mu _r b)^2} \frac{2\eta +1}{(1+\eta)^2} 
\frac{\langle~2u~ \rangle}{\langle~\upsilon ^2~\rangle}]  
\label{2.12}
\eeq 
We should remark that even though the quantity $\bar {\Sigma} _{spin}$
can be a function of u, in actual practice it is indepenent of u. The same
is true of te less important term $\bar{\Sigma} _{V}$
In the above expressions $F(u)$ is the nuclear form factor and
\beq
F_{\rho \rho^{\prime}}(u) =  \sum_{\lambda,\kappa}
\frac{\Omega^{(\lambda,\kappa)}_\rho( u)}{\Omega_\rho (0)} \,
\frac{\Omega^{(\lambda,\kappa)}_{\rho^{\prime}}( u)}
{\Omega_{\rho^{\prime}}(0)} 
, \qquad \rho, \rho^{\prime} = 0,1
\label{2.11} 
\eeq
are the spin form factors with
\beq
u = q^2b^2/2
\label{2.15} 
\eeq
b being the harmonic oscillator size parameter and q the momentum trasfer
to the nucleus.
The quantity u is also related to the experimentally measurable energy transfer
 Q via the relations
\beq
Q=Q_{0}u, \qquad Q_{0} = \frac{1}{A m_{N} b^2} 
\label{2.16b} 
\eeq
 The detection rate for a particle with velocity ${\boldmath \upsilon}$ 
and a target with mass $m$ detecting in the direction $\bf {e}$ will be 
denoted by $R(\rightarrow \bf {e})$. Then one defines the undirectiona rate 
$ R_{undir}$ via the equations 
via the equations 
\beq
R_{undir} =\frac{dN}{dt} =\frac{\rho (0)}{m_{\chi}} \frac{m}{A m_N} 
\sigma (u,\upsilon)
[\mid {\boldmath \upsilon}.\hat{e}_x\mid +  
\mid {\boldmath \upsilon}.\hat{e}_y\mid +  
\mid {\boldmath \upsilon}.\hat{e}_z\mid ]  
\label{2.17} 
\eeq
 $\rho (0) = 0.3 GeV/cm^3$ is the LSP density in our vicinity. 
This density has to be consistent with the LSP velocity distribution 
(see next section).

The differential undirectional  rate can be written as
\beq
dR_{undir} = \frac{\rho (0)}{m_{\chi}} \frac{m}{A m_N} 
d\sigma (u,\upsilon)
[\mid {\boldmath \upsilon}.\hat{e}_x\mid  +  
\mid {\boldmath \upsilon}.\hat{e}_y\mid  +  
\mid {\boldmath \upsilon}.\hat{e}_z\mid  ]  
\label{2.18}  
\eeq
where $d\sigma(u,\upsilon )$ is given by Eq. ( \ref{2.9})

 The directional rate in the direction $\hat{e}$ takes the form:
\beq
R_{dir} =R(\rightarrow~e)-R(\rightarrow~-e) =\frac{\rho (0)}{m_{\chi}} 
         \frac{m}{A m_N} 
{\boldmath \upsilon.e} ~  
\sigma (u,\upsilon)
\label{2.19}  
\eeq
and the corresponding differential rate is given by
\beq
dR_{dir} = \frac{\rho (0)}{m_{\chi}} \frac{m}{A m_N} 
{\boldmath \upsilon.e} ~  
d\sigma (u,\upsilon)
\label{2.20}  
\eeq

\section{Convolution of the Event Rate}
 
 We have seen that the event rate for LSP-nucleus scattering depends on the
relative LSP-target velocity. In this section we will examine the consequences 
of the earth's
revolution around the sun (the effect of its rotation around its axis is
expected to
be negligible) i.e. the modulation effect. This can be accomplished by
convoluting the rate with the LSP velocity distribution.Hitherto
such a consistent choice can be a Maxwell distribution ~\cite{Jungm}
\beq
f(\upsilon^{\prime}) = (\sqrt{\pi}\upsilon_0)^{-3}  e^{-(\upsilon^{\prime}/\upsilon_0)^2}
\label{3.1}  
\eeq
\beq
v_0 = \sqrt{(2/3) \langle v^2 \rangle } =220 Km /s
\label{3.2}  
\eeq
i.e. ${ v}_0$ is the velocity of the sun around the center of the galaxy.
In the present paper following the work of Drukier, see Ref. ~\cite{Druk},
 we will assume that the velocity distribution is
only axially symmetric, i.e. of the form
\beq
f(\upsilon^{\prime},\lambda) =  N(y_{esc},\lambda)( \sqrt{\pi}\upsilon_0)^{-3}) 
                              [ f_1(\upsilon^{\prime},\lambda)-
                                f_2(\upsilon^{\prime},\upsilon_{esc},\lambda)]
\label{3.3}  
\eeq
with
\beq
f_1(\upsilon^{\prime},\lambda)=exp[(- \frac{(\upsilon^{\prime}_x)^2+
                             (1+\lambda)((\upsilon^{\prime}_y)^2 +
                              (\upsilon^{\prime}_z)^2)}{\upsilon^2_0}]
\label{3.4}  
\eeq
\beq
f_2(\upsilon^{\prime},\upsilon_{esc},\lambda)=exp[- \frac{\upsilon^2_{esc}+
                                             \lambda((\upsilon^{\prime}_y)^2 +
                              (\upsilon^{\prime}_z)^2)}{\upsilon^2_0}]
\label{3.5}  
\eeq
where  $\upsilon_{esc}$ is the escape velocity in the
gravitational field of the galaxy, $\upsilon_{esc}=625Km/s$ \cite {Druk}.
In the above expressions $\lambda$ is a parameter, which describes the asymmetry 
and takes values between 0 and 1 and  N is a proper normalization constant given
by
\barr
\frac{1}{N(\lambda,y_{esc})}  & = & \frac{1}{\lambda +1}[erf(y_{esc})-
           e^{-(\lambda+1)y^2_{esc}}~\frac{erf(i~\sqrt{\lambda}~y_{esc})}
             {i~\sqrt{\lambda}}
\nonumber \\ 
 &-&\frac{e^{-y^2_{esc}}}{\lambda}
           [\frac{2}{\sqrt{\pi}}~y_{esc}-e^{-\lambda~y^2_{esc}}~
   \frac{erf(i~\sqrt{\lambda}~y_{esc})} {i~\sqrt{\lambda}}]
\label{3.5a}  
\earr
with $y_{esc}=\frac{\upsilon_{esc}}{\upsilon_0}$ and erf(x) the error function
given by
\beq
erf(x) =\frac{2}{\sqrt{\pi}} \int^{x}_{0} dt e^{-t^2}
\label{3.5b}  
\eeq
For $y_{esc} \rightarrow \infty$ we get the simple expression 
$N^{-1}=\lambda+1$

The z-axis is chosen in the direction of the disc's rotation, i.e. in the 
direction of the motion of the
the sun, the y-axis is perpendicular to the plane of the galaxy and the x-axis
is in the radial direction. 
In the above frame we find that the position of the axis of the ecliptic is 
determined by the angle $\gamma \approx 29.80$ (galactic latitude) and the
azimuthal angle $\omega = 186.3^0$ measured on the galactic plane
from the ${\bf \hat  z}$ axis~\cite{KVprd}.  Thus, the axis of the ecliptic 
lies very close to the $y,z$ plane and the velocity of the earth around the
sun is
\beq
{\bf \upsilon}_E \, = \, {\bf \upsilon}_0 \, + \, {\bf \upsilon}_1 \, 
= \, {\bf \upsilon}_0 + \upsilon_1(\, sin{\alpha} \, {\bf \hat x}
-cos {\alpha} \, cos{\gamma} \, {\bf \hat y}
+cos {\alpha} \, sin{\gamma} \, {\bf \hat z} \,)
\label{3.6}  
\eeq
where $\alpha$ is the phase of the earth's orbital motion, $\alpha =2\pi 
(t-t_1)/T_E$, where $t_1$ is around second of June and $T_E =1 year$.

We are now in a position to express the above distribution in the laboratory 
frame, i.e.
\beq
f({\bf \upsilon} \, , \,  \lambda \, , \, {\bf \upsilon}_E)=  
                  f_3(\upsilon,\upsilon_E,\lambda)- f_4(\upsilon,\upsilon_{esc},\lambda)
\label{3.7}  
\eeq
with
\barr
f_3(\upsilon,\upsilon_E,\lambda) & = &exp[- \frac{(\upsilon_x+ 
                         \upsilon_1 sin\alpha)^2 }{\upsilon^2_0}]
\nonumber \\ & \times &
            exp[-\frac{(1+\lambda)((\upsilon_y + \upsilon_1 cos\gamma 
                      sin\alpha)^2+( \upsilon_z+\upsilon_0+\upsilon_1
                      sin\gamma cos\alpha)^2)}{\upsilon^2_0}]
\nonumber \\
\label{3.8}  
\earr
\beq
f_4(\upsilon,\upsilon_{esc},\lambda)=exp[- \frac{\upsilon^2_{esc}+
                        \lambda((\upsilon_y + \upsilon_1 cos\gamma 
                      sin\alpha)^2+( \upsilon_z+\upsilon_0+\upsilon_1
                      sin\gamma cos\alpha)^2)}{\upsilon^2_0}]
\label{3.9}  
\eeq


\section{Expressions for the Differential Event Rate in the Presence
of Velocity Dispersion}
 We will begin with the undirectional  rate.
\subsection{Expressions for the Undirectional Differential Event Rate}

The mean value of the undirectional event rate of Eq. (\ref {2.19}), 
is given by
\beq
\Big<\frac{dR_{undir}}{du}\Big> =\frac{\rho (0)}{m_{\chi}} 
\frac{m}{A m_N} 
\int f({\bf \upsilon}, {\boldmath \upsilon}_E) 
[\mid {\boldmath \upsilon}.\hat{e}_x\mid  +  
\mid {\boldmath \upsilon}.\hat{e}_y\mid  +  
\mid {\boldmath \upsilon}.\hat{e}_z\mid ]  
                       \frac{d\sigma (u,\upsilon )}{du} d^3 {\boldmath \upsilon} 
\label{3.10} 
\eeq
From now on we will omit the subscript $undir$ in the case of the 
undirectional rate. The above expression can be more conveniently written as
\beq
\Big<\frac{dR}{du}\Big> =\frac{\rho (0)}{m_{\chi}} \frac{m}{Am_N} \sqrt{\langle
\upsilon^2\rangle } {\langle \frac{d\Sigma}{du}\rangle } 
\label{3.11}  
\eeq
where
\beq
\langle \frac{d\Sigma}{du}\rangle =\int
           \frac{   [\|{\boldmath \upsilon}.\hat{e}_x\| +  
\|{\boldmath \upsilon}.\hat{e}_y\| +  
\|{\boldmath \upsilon}.\hat{e}_z\| ]  }
{\sqrt{ \langle \upsilon^2 \rangle}} f({\boldmath \upsilon}, 
         {\boldmath \upsilon}_E)
                       \frac{d\sigma (u,\upsilon )}{du} d^3 {\boldmath \upsilon}
\label{3.12}  
\eeq
It is convenient to work in spherical coordinates. But even then the angular
sun is small. Thus introducing the parameter
\beq
\delta = \frac{2 \upsilon_1 }{\upsilon_0}\, = \, 0.27,
\label{3.13}  
\eeq
expanding in powers of $ \delta $ and keeping terms up to linear in it we can
manage to perform the $\phi$ integration using standard contour integral
techniques  and express the result in terms of the two modified
Bessel functions $I_m(\frac{\lambda \upsilon^2}{2 \upsilon^2_0}(1-t^2))$ with
$t=cos\theta$ and m=0,1. Thus the angular integration of Eq. \ref {3.7} yields
\barr
\tilde{M}_i(\lambda,y) & = & 2 \pi 
\nonumber \\       &\times&exp[-(y^2+1)(1+\lambda)] \tilde{\Lambda}_i(\lambda,y)- 
             exp[-(y^2_{esc}+ \lambda y^2)] \tilde{\Lambda}^{'}_i(\lambda,y),
 i=1,2
\nonumber\\
\label{3.14}  
\earr
where $\tilde{\Lambda}_i,\tilde{\Lambda}^{'}_i$ come from 
$f_3,f_4$ respectively. We find
\beq
\tilde{\Lambda}_1(\lambda,y)=  \tilde{\Lambda}_{1,1}(\lambda,y) 
                               +\tilde{\Lambda}_{1,2}(\lambda,y) 
                               +\tilde{\Lambda}_{1,3}(\lambda,y) 
\label{3.15}  
\eeq
with
\beq
\tilde{\Lambda}_{1,3}(\lambda,y)=  \int_{-1}^{1} d t 
exp[(\zeta^2/2-2(\lambda+1)y t)]\mid t\mid I_0(\zeta^2/2)
\label{3.16}  
\eeq
\beq
\tilde{\Lambda}_{1,2}(\lambda,y)= \frac{1}{\sqrt{\pi}}~ \int_{-1}^{1} d t 
exp[(-2(\lambda+1)y t)](1-t^2)^{1/2}\frac{erf(i~\zeta)}{i~\zeta}
\label{3.17}  
\eeq
\beq
\tilde{\Lambda}_{1,1}(\lambda,y)= \frac{1}{\sqrt{\pi}}~ \int_{-1}^{1} d t 
               exp[(-2(\lambda+1)y t)](1-t^2)^{1/2}
               \frac{exp(\zeta^2) ~erf(\zeta)}{\zeta}
\label{3.18}  
\eeq
where $\zeta~=~y~(\lambda~(1-t^2))^{1/2}$ and the function $erf(i\zeta)$ given 
by Eq. \ref {3.5b}. Furthermore 
\beq
\tilde {\Lambda}_2(\lambda,y)=- (\lambda+1) \tilde {\Lambda}^{''}_1(\lambda,y)
\label{3.19}  
\eeq
where $\tilde {\Lambda}^{''}_1(\lambda,y)$ is obtained from 
$\tilde{\Lambda}_1(\lambda,y)$ by
adding in the integrands the extra factor $ty+1$. We should mention that in
the last integral we have ommitted the numerical factor 
$\delta cos\alpha sin\gamma$.  Note that in the case 
$\tilde{\Lambda}_2(\lambda,y)$,
which is asociated with the modulation amplitude, only the z-component of
the velocity in the exponential contributes. Hence the dependence on the
earth's phase is $cos \alpha$.
 The formulas for the second term in Eq. \ref {3.14} for 
$\tilde{\Lambda}^{'}_i(\lambda,y)$ are obtained by   a mere replacement
of the expression $\lambda+1$ by $\lambda$. In all the above expressions
$y=(\upsilon / \upsilon_0)$ (not to be confused with the y-coordinate).

It is convenient to separate out the asymmetric contribution from the usual
one by writing
\beq
 2 y^2 \tilde{\Lambda}_1(\lambda,y) =\tilde{F}_0(\lambda,(\lambda+1)2y)
 +\tilde{G}_0(\lambda,y)
\label { 3.20a}  
\eeq
\beq
2 y^2 \tilde{\Lambda}_2(\lambda,y) =\tilde{F}_1(\lambda,(\lambda+1)2y)
+\tilde{G}_1(\lambda,y)
\label { 3.20b}  
\eeq
The functions $\tilde{F}_i$ have been obtained by considering the leading non
vanishing term  in the zeta expansion of the integrands of the expressions 
(\ref{3.16})-(\ref{3.18}). 
Thus 
\beq
\tilde{G}_0(0,y) =0~~~~~ ,~~~~~\tilde{G}_1(0,y)=0
\label { 3.20c}  
\eeq
\beq
\tilde{F}_0(\lambda,x) = (\lambda+1)^{-2}[x~sinh(x)-cos(x)+1+x~I_{1}(x)]  
\label{3.21a}  
\eeq
\barr
\tilde{F}_1(\lambda,x) & = &  (1+\lambda)^{-2}~
   [(2+\lambda)((x^2/(2(2+\lambda))+1)~cosh(x)- x~sinh(x)-1)
\nonumber\\
 &+& x^{2}~I_{2}[x]-(\lambda+1)~x~I_{1}(x)]  
\label{3.21b}  
\earr
note that here $x=(\lambda+1)2y$. $I_m(x)$ is the modified bessel function of
order m.
The funcions $\tilde{G}$ cannot bo obtained analytically, but they can easily 
be expressed as a rapidly convergent series in $y=\frac{
\upsilon}{\upsilon_0}$, which will not be given here.  

 Similarly
\beq
 \tilde{G}^{'}_i(\lambda,y) =
 2 y^2 \tilde{\Lambda}^{'}_i(\lambda,y)~~ ,~~ i=1,2
\label { 3.22}  
\eeq

Thus the folded non-directional event rate takes the form
\beq
\langle \frac{d\Sigma}{du} \rangle  = 
                          \bar {\Sigma} _{S} \bar {F}_0(u) +
        \frac{\langle \upsilon ^2 \rangle}{c^2}\bar {\Sigma} _{V} \bar {F}_1(u) 
                          +\bar {\Sigma} _{spin} \bar {F}_{spin}(u) 
\label{3.23}  
\eeq
where the $\bar{\Sigma} _{i},i=S,V,spin$ are given by Eqs. (\ref {2.10})-
(\ref {2.12}). 

The quantities 
$\bar{F}_0,\bar{F}_1,\bar{F}_{spin}$  
are obtained from the corresponding form factors via the equations
\beq
\bar{F}_{k}(u) = F^2(u)\Psi_k(u) \frac{(1+k)a^2 }{2k+1}~~,~~ k = 0,1
\label{3.24}  
\eeq
\beq
\bar{F}_{spin}(u) = F_{11}(u)\Psi_0(u)a^2 
\label{3.24a}  
\eeq
\beq
\label{3.25}  
\eeq
\beq
\tilde{\Psi}_k(u) =  [\tilde {\psi} _{(0),k}(a\sqrt{u})+
            0.135 \cos \alpha \tilde{\psi} _{(1),k}(a\sqrt{u})]  
\label{3.26}  
\ee
with
\beq
a = \frac{1}{\sqrt{2} \mu _rb\upsilon _0}  
\label{3.27}  
\eeq
and
\beq
\tilde{\psi} _{(l),k}(x)= N(y_{esc},\lambda) e^{-\lambda}(e^{-1} 
   \tilde{\Phi}_{(l),k}(x)-
                  exp[-y^2_{esc}] \tilde{\Phi}^{'}_{(l),k}(x))  
\label{3.28}  
\eeq
\beq
\tilde {\Phi}_{(l),k}(x) =  \frac{2}{\sqrt{6 \pi}} \int_x^{y_{esc}} dy y^{2k-1}
              exp {(-(1+\lambda) y^2)})( \tilde{F}_{l}(\lambda,(\lambda+1) 2 y )
+\tilde{G}_{l}(\lambda,y)))
\label{3.29}  
\eeq
\beq
\tilde{\Phi}^{'}_{(l)_,k}(x) =  \frac{2}{\sqrt{6 \pi}} \int_x^{y_{esc}} dy y^{2k-1}
                     exp {(-\lambda y^2)})\tilde {G}^{'}_{l}(\lambda,y))
\label{3.30}  
\eeq
The undirectional differential rate takes the form
\beq
\langle \frac{dR}{du} \rangle  = \bar{R}  t T(u) [(1 + \cos \alpha H(u))] 
\label{3.31}  
\eeq
In the above expressions $\bar{R}$ is the rate obtained in the conventional 
approach \cite {JDV} by neglecting the folding with the LSP velocity and the
momentum transfer dependence of the differential cross section, i.e. by
\beq
\bar{R} =\frac{\rho (0)}{m_{\chi}} \frac{m}{Am_N} \sqrt{\langle
v^2\rangle } [\bar{\Sigma}_{S}+ \bar{\Sigma} _{spin} + 
\frac{\langle \upsilon ^2 \rangle}{c^2} \bar{\Sigma} _{V}]
\label{3.39b}  
\eeq
where $\bar{\Sigma} _{i}, i=S,V,spin$ have been defined above, see Eqs
 (\ref {2.10}) - (\ref {2.12}). 

The factor 
$T(u)$ takes care of the u-dependence of the unmodulated differential rate. It
is defined so that
\beq
 \int_{u_{min}}^{u_{max}} du T(u)=1.
\label{3.30a}  
\eeq
i.e. it is the relative diffrential rate. $u_{min}$ is determined by the energy 
cutoff due to the performance of the detector. $u_{max}$ is determined by the 
escape velocity $\upsilon_{esc}$ via the relations:
\beq
u_{max}=\frac{y_{esc} ^2}{a^2}~~ 
\label{3.30b}  
\eeq
On the other hand
$H(u)$ gives the energy tranfer dependent modulation amplitude.
The quantity $t$ takes care of the modification of the total rate due to the
nuclear form factor and the folding with the LSP velocity distribution.
 Since the functions $\bar{F}_0(u),\bar{F}_1$ and $\bar{F}_{spin}$ in principle
have a different dependence on u, the funcions $T(u),H(u)$ and $t$ in priciple 
depend on the SUSY parameters. If, however, we ignore the small vector 
contribution and assume (i) the scalar and axial (spin) dependence on u is the
same or (ii) only one mechanism (S, V, spin) dominates the, parameter $\bar{R}$ 
contains the dependence on all SUSY parameters. The other factors depend only
on  the LSP mass and the nuclear parameters. 
 More specifically considering only the scalar interaction we get
 $\bar{R} \rightarrow \bar{R}_{S}$ and
\beq
t~ T(u) =  a^2 F^2(u) \tilde{\psi} _{(0),0}(a \sqrt{u})    
\label{3.48}  
\eeq

For the spin interaction we get a similar expression except that
$\bar{R}\rightarrow \bar{R}_{spin}$ and 
$F^2 \rightarrow F_{\rho ,\rho^{\prime}}$.
Finally for completeness we will consider the less important vector
contribution.
 We get $\bar{R} \rightarrow \bar{R}_{V}$ and
\beq
t~ T(u) =  F^2(u) [ \tilde{\psi} _{(0),1}(a \sqrt{u})
                       -\frac{1}{(2 \mu _r b)^2} \frac{2\eta +1}{(1+\eta)^2}
                        u\, \tilde{\psi} _{(0),0}(a \sqrt{u})] \frac{2a^2}{3}    
\label{3.52}  
\eeq
 The quantity $T(u)$ depends on nuclear physics through the form factors
or the spin response functions and the parameter $a$. The modulation amplitude
takes the form

\beq
H(u) =0.135\frac{\tilde{\psi}_{(1),k}(a \sqrt{u})}
                {\tilde{\psi} _{(0),k}(a \sqrt{u})},  
        l=1,3  
\label{3.45}  
\eeq 
Thus in this case the $H(u)$ depends only on $a\sqrt{u}$, which coincides with 
the parameter x of Ref. \cite{Smith}, i.e.
only on the momentum transfer, the reduced mass and the size of the nucleus.

 Returning to the differential rate it is sometimes convenient to use the
quantity $T(u) H(u)$ rather than $H$, since H(u) may appear artificially 
increasing function of u
due to the decrease of $T(u)$ ( in obtaining H(u) we have divided T(u)) 

Before concluding this subsection we should mention that the above angular 
integrations can also be done even if the velocity distribution is triaxial.
We will not explore this further since one has too many parameters.

\subsection{Expressions for the Directional Differential Event Rate}

The mean value of the directional differential event rate of Eq. (\ref {2.20}), 
is defined by
\beq
\Big<\frac{dR}{du}\Big>_{dir} =\frac{\rho (0)}{m_{\chi}} 
\frac{m}{A m_N} 
\int f({\bf \upsilon}, {\boldmath \upsilon}_E) {\boldmath \upsilon.e} 
                       \frac{d\sigma (u,\upsilon )}{du} d^3 {\boldmath \upsilon} 
\label{4.10} 
\eeq
where ${\bf \hat e}$ is the unit vector in the direction of observation. 
It can be more conveniently expressed as
\beq
\Big<\frac{dR}{du}\Big>_{dir} =\frac{\rho (0)}{m_{\chi}} \frac{m}{Am_N} \sqrt{\langle
\upsilon^2\rangle } {\langle \frac{d\Sigma}{du}\rangle }_{dir} 
\label{4.11}  
\eeq
where
\beq
\langle \frac{d\Sigma}{du}\rangle _{dir}=\int \frac{ {\boldmath \upsilon.e}} 
{\sqrt{ \langle \upsilon^2 \rangle}} f({\boldmath \upsilon}, {\boldmath \upsilon}_E)
                       \frac{d\sigma (u,\upsilon )}{du} d^3 {\boldmath \upsilon}
\label{4.12}  
\eeq

 Working as in the previous subsection, i.e by 
expanding in powers of $ \delta $ and keeping terms up to linear in it we can
manage to perform the $\phi$ integration using standard contour integral
techniques  and express the result in terms of the two modified
Bessel functions $I_m(\frac{\lambda \upsilon^2}{2 \upsilon^2_0}(1-t^2))$ with
$t=cos\theta$ and m=0,1. Thus the angular integration of Eq. \ref {3.7} yields
\barr
M_i(\lambda,y) & = & 2 \pi 
\nonumber \\       &\times&exp[-(y^2+1)(1+\lambda)] \Lambda_i(\lambda,y)- 
             exp[-(y^2_{esc}+ \lambda y^2)] \Lambda^{'}_i(\lambda,y), i=1,4
\nonumber\\
\label{4.14}  
\earr
where $\Lambda_i,\tilde{\Lambda}-i$ come from $f_3,f_4$ respectively and are
given by
\beq
\Lambda_1(\lambda,y)=  \int_{-1}^{1} d t 
exp[-((\lambda/2)y^2(1-t^2)+2(\lambda+1)y t)] t I_0((\lambda/2)y^2(1-t^2))
\label{4.15}  
\eeq
\beq
\Lambda_2(\lambda,y)=  \int_{-1}^{1} d t 
exp[-((\lambda/2)y^2(1-t^2)+2(\lambda+1)y t)] t^2 I_0((\lambda/2)y^2(1-t^2))
\label{4.16}  
\eeq
\beq
\Lambda_3(\lambda,y)=  \int_{-1}^{1} d t 
exp[-((\lambda/2)y^2(1-t^2)+2(\lambda+1)y t)] (1-t^2) I_0((\lambda/2)y^2(1-t^2))
\label{4.17}  
\eeq
\beq
\Lambda_4(\lambda,y)=  \int_{-1}^{1} d t 
exp[-((\lambda/2)y^2(1-t^2)+2(\lambda+1)y t)] (1-t^2) I_1((\lambda/2)y^2(1-t^2))
\label{4.18}  
\eeq
and analogous expressions for $\Lambda^{'}$ with the mere replacement
of the expression $\lambda+1$ by $\lambda$. Again in  the above expressions
y=$(\upsilon / \upsilon_0)$ (not to be confused with the y-coordinate).
The above integrals can be expressed in terms of hypergeometric functions as 
follows

\beq
\Lambda_1(\lambda,y)= - \sum^{\infty}_{k=0} 
                \frac{2(2(\lambda+1)y)^{(2k+1)}}{(2k+3)((2k+1)!)} 
                F^1_1(\frac{1}{2},\frac{2k+5}{2},\lambda y^2)
\label{4.19}  
\eeq
\beq
\Lambda_2(\lambda,y)= (\lambda+1) \sum^{\infty}_{k=0} 
                \frac{2(2(\lambda+1)y)^{(2k)}}{(2k+3)((2k)!)} 
                F^1_1(\frac{1}{2},\frac{2k+5}{2},\lambda y^2)
\label{4.20}  
\eeq
\beq
\Lambda_3(\lambda,y)= (\lambda+1) \sum^{\infty}_{k=0} 
                \frac{2(2(\lambda+1)y)^{(2k)}}{(2k+1)(2k+3)((2k)!)} 
                F^2_2(\frac{1}{2},2,1,\frac{2k+5}{2},\lambda y^2)
\label{4.21}  
\eeq
\beq
\Lambda_4(\lambda,y)= (\lambda+1)\lambda  y^2 \sum^{\infty}_{k=0} 
                \frac{2(2(\lambda+1)y)^{(2k)}}{(2k+1)(2k+3)(2k+5)((2k)!)} 
                F^1_1(\frac{3}{2},\frac{2k+7}{2},\lambda y^2)
\label{4.22}  
\eeq
\beq
\Lambda^{'}_1(\lambda,y)= - \sum^{\infty}_{k=0} 
                \frac{2(2\lambda y)^{(2k+1)}}{(2k+3)((2k+1)!)} 
                F^1_1(\frac{1}{2},\frac{2k+5}{2},\lambda y^2)
\label{4.23}  
\eeq
\beq
\Lambda^{'}_2(\lambda,y)=  \lambda \sum^{\infty}_{k=0} 
                \frac{2(2\lambda y)^{(2k)}}{(2k+3)((2k)!)} 
                F^1_1(\frac{1}{2},\frac{2k+5}{2},\lambda y^2)
\label{4.24}  
\eeq
\beq
\Lambda^{'}_3(\lambda,y)=  \lambda \sum^{\infty}_{k=0} 
                \frac{2(2\lambda y)^{(2k)}}{(2k+1)(2k+3)((2k)!)} 
                F^2_2(\frac{1}{2},2,1,\frac{2k+5}{2},\lambda y^2)
\label{4.25}  
\eeq
\beq
\tilde{\Lambda}_4(\lambda,y)=  (\lambda y)^2 \sum^{\infty}_{k=0} 
                \frac{2(2\lambda y)^{(2k)}}{(2k+1)(2k+3)(2k+5)((2k)!)} 
                F^1_1(\frac{3}{2},\frac{2k+7}{2},\lambda y^2)
\label{4.26}  
\eeq
 It is more convenient to define the functions $F_i$ ang $G_i$, $i=0,4$ as
follows
\beq
-2 y^2 \Lambda_1(\lambda,y)= F_0(2(\lambda+1)y) + G_0(\lambda,y)  
\label{4.27}  
\eeq
\beq
2 y^2 (\Lambda_1(\lambda,y)+ y \Lambda_2(\lambda,y))=F_1(\lambda,2(\lambda+1)) 
                                                    +G_1(\lambda,y)
\label{4.28}  
\eeq
\beq
4 y^3 (\Lambda_3(\lambda,y)- \Lambda_4(\lambda,y))=F_2(2(\lambda+1)) 
                                                  +G_2(\lambda,y)
\label{4.29}  
\eeq
\beq
4 y^3 (\Lambda_3(\lambda,y)+ \Lambda_4(\lambda,y))=F_3(2(\lambda+1)) 
                                                  +G_3(\lambda,y)
\label{4.30}
\eeq
 The functions $F_i$  are obtained by keeping the leading terms in the
expansion of the confluent hypergeommetric functions of Eqs. (\ref{4.19})-
(\ref{4.22}). 
Thus we find 
\beq
F_i(\chi) =\chi cosh \chi - sinh \chi ~~,~~i=0,2,3 
\label{4.31}  
\eeq
\beq
F_1(\lambda,\chi) \, = \, 2(1-\lambda)\, \Big[ \,(\frac{(\lambda+1)\chi^2}
{4(1-\lambda)} + 1) sinh\, \chi - 
\chi \,cosh \,\chi  \, \Big]
\label{4.32}
\eeq
 The purely asymmetric quantities $G_i$ satisfy
\beq
G_i(0,y)=0,~i=0,4  
\label{4.33}  
\eeq
They are expressed as rapidly convergent series in $y$, but they are not going
to be given here.
Similarly we define the functions $G^{'}$ via the equations
\beq
G^{'}_0(\lambda,y)=-2 y^2 \Lambda^{'}_1(\lambda,y)  
\label{4.35}  
\eeq
\beq
G^{'}_1(\lambda,y)=2 y^2 (\Lambda^{'}_1(\lambda,y)+ 
                                         y \Lambda^{'}_2(\lambda,y))  
\label{4.36}  
\eeq
\beq
G_2(\lambda,y)=4 y^3 (\Lambda^{'}_3(\lambda,y)- \Lambda{'}_4(\lambda,y))  
\label{4.37}  
\eeq
\beq
G^{'}_3(\lambda,y)=0 
\label{4.38}  
\eeq 
Thus the folded directional event rate takes the form
\beq
\langle \frac{d\Sigma}{du} \rangle_{dir}  = \frac{1}{2}~a^2
                          [\bar {\Sigma} _{S} F_0(u) +
                     \frac{\langle \upsilon ^2 \rangle}{c^2}
                          \bar {\Sigma} _{V} F_1(u) +
                          \bar {\Sigma} _{spin} F_{spin}(u) ]
\label{4.39}  
\eeq
where the $\bar{\Sigma} _{i},i=S,V,spin$ are given by Eqs. (\ref {2.10})-
(\ref {2.12}). 
The quantities 
$F_0,F_1,F_{spin}$  
are now obtained from the corresponding form factors via the equations
\beq
F_k(u) = F^2(u)\Psi_k(u)\frac{(1+k)a^2}{2k+1} , k = 0,1
\label{4.40}  
\eeq
\beq
F_{spin}(u) = F_{11}(u) \Psi_0(u) a^2
\label{4.41}  
\eeq
\barr
\Psi_k(u)& = &\frac{1}{2} [(\psi _{(0),k}(a\sqrt{u})+
            0.135 \cos \alpha \psi _{(1),k}(a\sqrt{u})){\bf e}_z.{\bf e}  
\nonumber \\ & - &
            0.117 \cos \alpha \psi _{(2),k}(a\sqrt{u}) {\bf e}_y.{\bf e}+  
            0.135 \sin \alpha \psi _{(3),k}(a\sqrt{u})  {\bf e}_x.{\bf e}]  
\label{4.42}  
\earr
with
\beq
\psi _{(l),k}(x)= N(y_{esc},\lambda) e^{-\lambda}(e^{-1} \Phi_{(l),k}(x)-
                  exp[-y^2_{esc}] \Phi{'}_{(l),k}(x))  
\label{3.36}  
\eeq
\beq
\Phi_{(l),k}(x) =  \frac{2}{\sqrt{6 \pi}} \int_x^{y_{esc}} dy y^{2k-1}
                     exp {(-(1+\lambda) y^2)})( F_{l}(2 y )+G_{l}(\lambda,y)))
\label{3.37}  
\eeq
\beq
\Phi^{'}_{(l),k}(x) =  \frac{2}{\sqrt{6 \pi}} \int_x^{y_{esc}} dy y^{2k-1}
                     exp {(-\lambda y^2)}) G{'}_{l}(\lambda,y))
\label{3.38}  
\eeq
 If we consider each mode (scalar, spin vector) separately the directional
rate takes the form
\beq
\langle \frac{dR}{du} \rangle_{dir}  = \frac{\bar{R}}{2}~t^0~R^0 
                       [(1 + \cos \alpha H_1(u)){\bf e}_z.{\bf e}- 
                       \cos \alpha H_2(u) {\bf e}_y.{\bf e}+  
                           \sin \alpha H_3(u)  {\bf e}_x.{\bf e}]  
\label{3.40}  
\eeq

In other words the non directional differential modulated amplitude is described
in terms of the three parameters, $H_l(u)$, l=1,2 and 3. The unmodulated one is
$R^0(u)$, which is again normalized to unity. It is the relative differential 
rate, i.e. the
differential rate divided by the total rate, in the absence of modulation, i.e.
The parameter $t^0$ entering Eq. (\ref {3.40}) takes care of whatever 
modifications are needed due to the convolution of the non modulated total rate
with the 
LSP velocity  distribution in the presence of the nuclear form factors. 

From Eqs. (\ref {4.38}) - (\ref {3.40}) we see that if we consider each
mode separately the differential modulation amplitudes $H(l)$ take the form
\beq
H_l(u) =0.135\frac{\psi^{(l)}_{k}(a \sqrt{u})}{\psi ^{(0)}_{k}(a \sqrt{u})}~~,
         ~~l=1,3~~;~~  
H_2(u) =0.117 \frac{\psi^{(2)}_{k}(a \sqrt{u})}{\psi ^{(0)}_{k}(a \sqrt{u})}    
\label{4.45}  
\eeq 
Thus in this case the $H_l$ depend only on $a\sqrt{u}$, which coincides with 
the parameter x of Ref. \cite{Smith}.
 We note that in
the case $\lambda=0$ we have $H_2=0.117$ and $H_3=0.135$, so that there remains
This means that, if we neglect the coherent vector contribution, which,
as we have mentioned, is small, $H_l$ essentially depends 
only on the momentum transfer, the reduced mass and the size of the nucleus.

 Returning to the differential rate it is sometimes convenient to use the
quantity $R_l$ rather than $H_l$ defined by
\beq
R_l = R^0 H_l, \,  l=1,2,3.
\label{3.46}  
\eeq
The reason is that $H_l$, being the ratio of two quantities, may appear
superficially large due to the denominator becoming small.
once again if one mechanism dominates the parameters $R_0$ and $R_l$ 
are independent of the particular SUSY model considered, except the 
LSP mass. In fact we find for the scalar interaction we get
 $\bar{R} \rightarrow \bar{R}_{S}$ and
\beq
t^0~ R^0(u) =  a^2 F^2(u) \psi ^{(0)}_0(a \sqrt{u})    
\label{4.48}  
\eeq

For the spin interaction we get a similar expression except that
$\bar{R}\rightarrow \bar{R}_{spin}$ and 
$F^2 \rightarrow F_{\rho ,\rho^{\prime}}$.
Finally for completeness we will consider the less important vector
contribution.
 We get $\bar{R} \rightarrow \bar{R}_{V}$ and
\beq
t^0~ R^0(u) =  F^2(u) [ \psi ^{(0)}_1(a \sqrt{u})    
                       -\frac{1}{(2 \mu _r b)^2} \frac{2\eta +1}{(1+\eta)^2}
                        u\, \psi ^{(0)}_0(a \sqrt{u}) ] \frac{2a^2}{3}   
\label{4.52}  
\eeq
 The quantity $R_0$ depends on nuclear physics through the form factors
or the spin response functions.

 Equation (\ref {3.40}) deviates from the simple trigonometric expression.a
The dependence on the phase of the earth is complicated. If we imagine,
however, that one can sum up all three directonal rates, with  the inclusion of
$H_2$ and $H_3$, the maximum does not occur at $\alpha=0$, but at
$\alpha = \alpha_H$ with
\beq
\alpha_H = tan^{-1}[\frac{ H_3(u)} {H_1(u) + H_2(u)}]   
\label{3.41a}  
\eeq
and the modulation at this value of the phase of the earth takes the value
\beq
H_{max} = [(H_1 + H_2)^2 + H_3^2]^{\frac{1}{2}}
\label{3.41b}  
\eeq
 There exists one minimum at $\alpha= \pi$, i.e. around Dec.2 and takes the
 value
\beq
H_{min} = H_2-H_1 
\label{3.42}  
\eeq
Whenever $H_1>H_2$ there exist two more minima $\alpha= \pi/2$ and $3(\pi/2)$, 
equal to $H_3$ and two secondary maxima. In all cases considered in this work
$H_3>H_2-H_1$ so that the interesting quantities are given by Eqs(\ref {3.41b} 
- \ref {3.42}). In any case it is useful to know the difference between the
maximum and the minimum, which takes the form
\beq
H_m = [(H_1 + H_2)^2 + H_3^2]^{\frac{1}{2}}-Min( H_1 -H_2 , H_3)
\label{3.43}  
\eeq


\section{The Total Modulated  Event Rates}

 Once again we will distinguish two possibilities, namely the directional and 
the non directional case. Integrating Eq. (\ref {3.40}) we obtain for the total
undirectional rate
\beq
R =  \bar{R}\, t \, [(1 + h(a,Q_{min})cos{\alpha})] 
\label{3.55}  
\eeq
where $Q_{min}$ is the energy transfer cutoff imposed by the detector.
 The modulation of the non-directional total event rate can
be described in terms of the parameter $h$. 

 The effect of folding
with LSP velocity on the total rate is taken into account via the quantity
$t$. The SUSY parameters have been absorbed in $\bar{R}$. From our 
discussion in the case of differential rate it is clear that strictly
speaking the quantities $t$ and $h$ also depend on the SUSY parameters. They do 
not depend on them, however, if one considers the scalar,spin etc. modes 
separately. 

 Let us now examine the directional rate.  Integrating Eq. (\ref {4.40}) we 
obtain
\barr
R_{dir}& = &  \bar{R}\, (t^0/2) \, 
                       [(1 + h_1(a,Q_{min})cos{\alpha}) {\bf e}~_z.{\bf e}
\nonumber\\  &-& h_2(a,Q_{min})\, 
cos{\alpha}{\bf e}~_y.{\bf e}
                      + h_3(a,Q_{min})\, 
sin{\alpha}{\bf e}~_x.{\bf e}]
\label{4.55}  
\earr
 Furthermore if we somehow manage to measure the directional rate in all 
directions we obtain:
\beq
R_{dir,all} = \bar{R}\, (t^0/2) \, [1 + h_1(a,Q_{min})\, cos \alpha +
            h_2(a,Q_{min})\, |cos{\alpha}| + 
            h_3(a,Q_{min})\, |sin{\alpha}|]
\label{4.56}  
\eeq
 We see that the modulation of the directional total event rate can
be described in terms of three parameters $h_l$ l=1,2,3. 
 In the special case of $\lambda=0$ we essentially have  one 
parameter, namely $h_1$, since then we have $h_2=0.117$ and $h_3=0.135$.

 The effect of folding
with LSP velocity on the total rate is taken into account via the quantity
$t^0$. All other SUSY parameters have been absorbed in $\bar{R}$, under the
same asumptions discussed above in the case of undirectional rates.

Given the functions$h_l(a,Q_{min})$ one can plot the the expression in
Eq. (\ref {4.56}) as a function of the phase of the earth $\alpha$. For
a gross description one can follow the procedure outlined above making 
the substitution $H \rightarrow h$. Thus the maximum occurs at $\pm \alpha_h$
with
\beq
\alpha_h = tan^{-1}[\frac{ h_3(a,Q_{min})}
 {h_1(a,Q_{min}) + h_2(a,Q_{min})}]   
\label{3.57}  
\eeq
The difference between the the maximum and the minimum is now given by
\beq
h_m = [(h_1 + h_2)^2 + h_3^2]^{\frac{1}{2}} - Min (h_1 - h_2,h_3)|
\label{3.58}  
\eeq
In all cases considered here $h_3>|h_1-h_2|$


\section{Discussion of the Results}
We have calculated the differential as well as the total event rates 
(directional and non directional) for elastic LSP-nucleus scattering for
the target $^{127}I$, including realistic form factors. Only the coherent mode
due to the scalar interaction was considered. The spin contribution will appear
elsewhere. Special attention was paid to the modulation effect
due to the annual motion of the earth. To this end we included not only 
the component of the earth's velocity in  
the direction of the sun's motion, as it has been done so far, but all of its
components. In addition both spherically
symmetric \cite {Jungm} as well only axially symmetric \cite {Druk}
LSP velocity distributions were examined. Furthermor we considered
the effects of the detector energy cutoffs, by studying two typical
cases $Q_{min}=10$ and 20 KeV both on the modulated and the unmodulated
aplitudes. We focused our attention on those aspects which do not depend on
the parameters of supersymmetry other then the LSP mass.

 The parameter $\bar {R}$, normally calculated in SUSY theories, was not
considered in this work. The interested reader is referred to the literature
\cite {ref1} $^,$ \cite {ref4} and, in our notation, to our previous work 
\cite {JDV} $^,$ \cite {KVprd} $^,$\cite {KVdubna}.\\

\subsection{The Undirectional Rates}

 Let us begin with the total rates, i.e. the quantities $t$ and $h$. In
Table 1 we show the dependence of $h$
on the components of the earth's velocity, in the symmetric case($\lambda=0$).
We see that the modulation 
amplitude increases for about $50~\%$ when all components of the earth's
velocity are included. 

In Table 2 we show how the quantities $t$ and $h$ depend on the detector energy
cutoff and the LSP mass for the symmetric case. In tables 3 and 4 we show the 
same quantities for $\lambda=0.5$ and $\lambda=1.0$ respectively. From these 
tables we see a dramatic incease of the modulation when the realistic axially 
symmetric velocity distribution is turned on. This means that the modulation 
amplitude can be exploited by the experimentalists. We further notice
that the modulation amplitude increases somewhat with cutoff enery.
This is due to the fact that the modulation amplitude decreases less rapidly
with the cutoff energy $Q_{min}$ than the unmodulated amplitude. This effect 
may be of use to the experimentalists, even though it occurs at the expense
of the total rate. 
\bigskip
\bigskip
\begin{table}[t]  
\caption{
 The dependence of the modualtion amplitude $h$ on the velocity of
the earth in the symmetric case $\lambda=0.$ and $Q_{min}=0.$}
\begin{center}
\footnotesize
\begin{tabular}{|l|crrrrrr|}
\hline
\hline
& & & & & & &     \\
&   \multicolumn{7}{c|}{LSP \hspace {.2cm} mass \hspace {.2cm} in GeV}  \\ 
\hline 
& &  & & & & &     \\
Velocity &   10   & 30  & 50  & 80  & 100 & 125 & 250  \\
\hline 
& & & & &  & &    \\
z-com&0.0453& 0.0320& 0.0179& 0.0075& 0.0041& 0.0015& -0.0033\\
all&0.0723& 0.0558& 0.0383& 0.0252& 0.0208& 0.0173& 0.0112\\
\hline
\hline
\end{tabular}
\end{center}
\end{table}
\bigskip
\bigskip
\begin{table}[t]  
\caption{The quantities $t$ and $h$ for $\lambda=0$ in the case of the
target $_{53}I^{127}$ for various LSP masses and $Q_{min}$ in KeV (for
definitions see text). Only the scalar contribution is considered.}
\begin{center}
\footnotesize
\begin{tabular}{|l|c|rrrrrrr|}
\hline
\hline
& & & & & & & &     \\
&  & \multicolumn{7}{|c|}{LSP \hspace {.2cm} mass \hspace {.2cm} in GeV}  \\ 
\hline 
& & & & & & & &     \\
Quantity &  $Q_{min}$  & 10  & 30  & 50  & 80 & 100 & 125 & 250   \\
\hline 
& & & & & & & &     \\

t&0.0&1.599& 1.134& 0.765& 0.491& 0.399& 0.328& 0.198\\
h&0.0&0.072& 0.056& 0.038& 0.025& 0.021& 0.017& 0.011\\
\hline 
& & & & & & & &     \\
t&10.&0.000& 0.276& 0.307& 0.236& 0.200& 0.170& 0.108\\
h&10.&0.000& 0.055& 0.028& 0.014& 0.010& 0.007& 0.001\\
\hline 
& & & & & & & &     \\
t&20.&0.000& 0.058& 0.117& 0.110& 0.098& 0.086& 0.058\\
h&20.&0.000& 0.084& 0.044& 0.024& 0.017& 0.013& 0.005\\
\hline
\hline
\end{tabular}
\end{center}
\end{table}
\bigskip
\bigskip
\begin{table}[t]  
\caption{The same quanitities with table I for $\lambda=0.5$}
\begin{center}
\footnotesize
\begin{tabular}{|l|c|rrrrrrr|}
\hline
\hline
& & & & & & & &     \\
&  & \multicolumn{7}{|c|}{LSP \hspace {.2cm} mass \hspace {.2cm} in GeV}  \\ 
\hline 
& & & & & & & &     \\
Quantity &  $Q_{min}$  & 10  & 30  & 50  & 80 & 100 & 125 & 250   \\
\hline 
& & & & & & & &     \\
t&0.0&1.690& 1.241& 0.861& 0.558& 0.453& 0.372& 0.224\\
h&0.0&0.198& 0.151& 0.107& 0.083& 0.076& 0.071& 0.063\\
\hline
t&10.&0.000& 0.267& 0.337& 0.268& 0.229& 0.194& 0.122\\
h&10.&0.000& 0.344& 0.175& 0.113& 0.097& 0.087& 0.072\\
\hline
h&20.&0.000& 0.000& 0.121& 0.123& 0.111& 0.098& 0.066\\
h&20.&0.000& 0.000& 0.267& 0.150& 0.124& 0.106& 0.081\\
\hline
\hline
\end{tabular}
\end{center}
\end{table}
\bigskip
\bigskip
\begin{table}[t]  
\caption{The same quanitities with table I for $\lambda=1.0$}
\begin{center}
\footnotesize
\begin{tabular}{|l|c|rrrrrrr|}
\hline
\hline
& & & & & & & &     \\
&  & \multicolumn{7}{|c|}{LSP \hspace {.2cm} mass \hspace {.2cm} in GeV}  \\ 
\hline 
& & & & & & & &     \\
Quantity &  $Q_{min}$  & 10  & 30  & 50  & 80 & 100 & 125 & 250   \\
\hline 
& & & & & & & &     \\
t&0.0&1.729& 1.299& 0.919& 0.600& 0.487& 0.399& 0.240\\
h&0.0&0.314&0.247& 0.181& 0.141& 0.131& 0.123& 0.112\\
\hline
t&10.&0.000& 0.252& 0.353& 0.289& 0.247& 0.209& 0.132\\
h&10.&0.000& 0.579& 0.291& 0.187& 0.163& 0.147& 0.124\\
\hline
t&20.&0.000& 0.000& 0.120& 0.131& 0.120& 0.106& 0.071\\
h&20.&0.000& 0.000& 0.455& 0.249& 0.205& 0.177& 0.137\\
\hline
\hline
\end{tabular}
\end{center}
\end{table}

 Let us now examine the differential rates, which are described by the functions
$T(u)$, $H(u)$ and $T(u) \times H(u)$. These are shown  for various
LSP masses and $Q_{min}$ in Fig. 1 ($\lambda=0.0$), Fig. 2
($\lambda=0.5$) and Fig. 3 ($\lambda=1.0$)
 We remind the reader that the dimensionless quantity u is related to the 
energy transfer Q 
via Eq. (\ref{2.17}) with $Q_0=60 KeV$ for $^{127}I$. 
The curves shown
correspond  to LSP masses as follows:
i) Solid line $\Longleftrightarrow m_{\chi}=30$ GeV.
ii) Dotted line $\Longleftrightarrow m_{\chi}=50$ GeV.
iii) Dashed line $\Longleftrightarrow m_{\chi}=80$ GeV.
v) Intermediate dashed line $\Longleftrightarrow m_{\chi}=100$ GeV.
vi) Fine solid line $\Longleftrightarrow m_{\chi}=125$ GeV.
vi) Long dashed line $\Longleftrightarrow m_{\chi}=250$ GeV.
If some curves of the above list seem to have been omitted, it is
understood that they fall on top of vi). Note that, due to our
normalization of T, the area under the corresponding curve is unity.
 This normalzation was adopted to bring the various graphs on scale 
since the absolute values may change
much faster as a function of the LSP mass.
\setlength{\unitlength}{1mm}
\begin{figure}
\begin{picture}(150,200)
\put(40,0){\epsfxsize=5cm \epsfbox{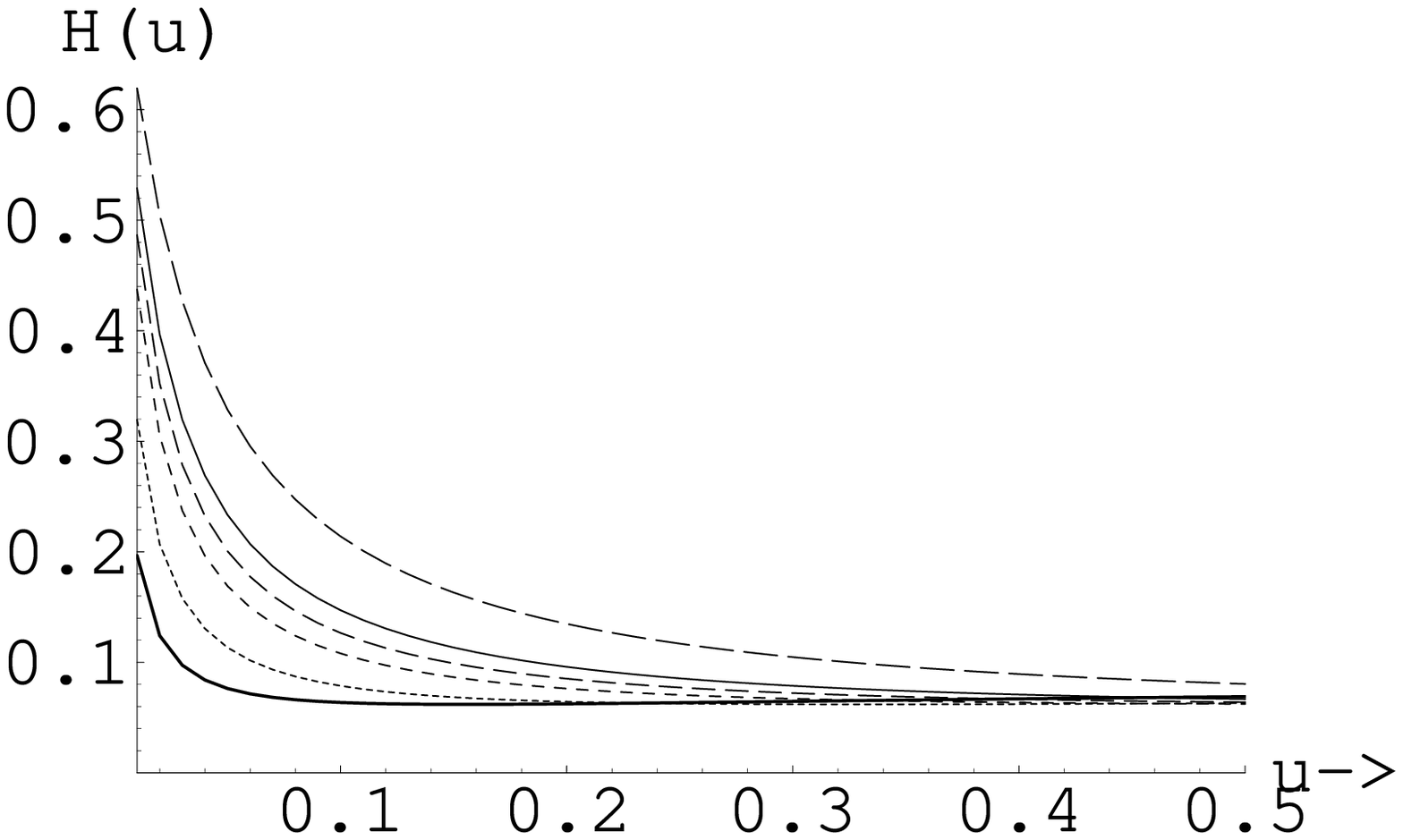}}
\put(40,0){\bf ($\lambda=0.0$, independent of $Q_{min}$)}
\put(10,60){\epsfxsize=5cm \epsfbox{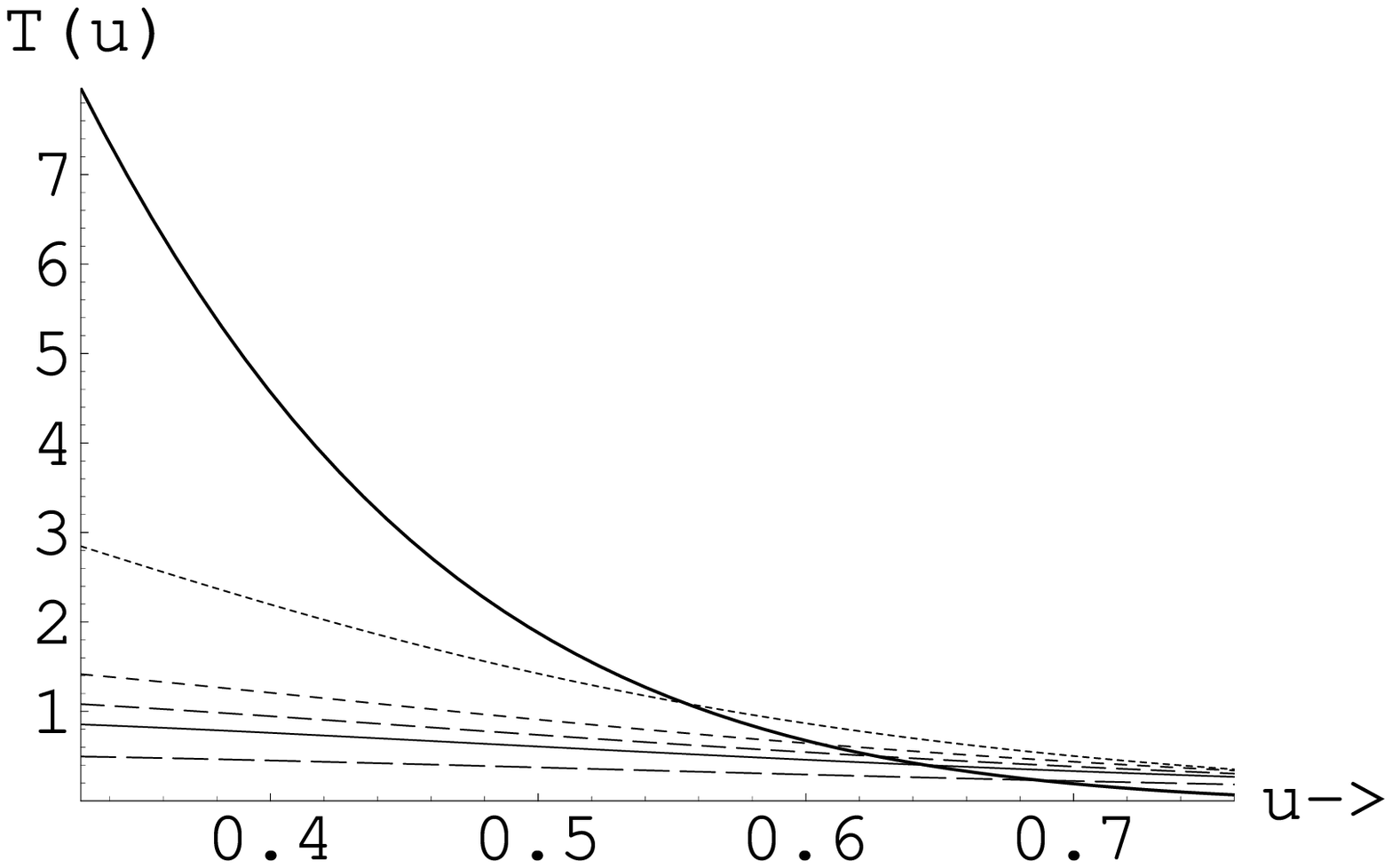}}
\put(10,60){\bf ($\lambda=0.0$, $Q_{min}=20.0$)}
\put(70,60){\epsfxsize=5cm \epsfbox{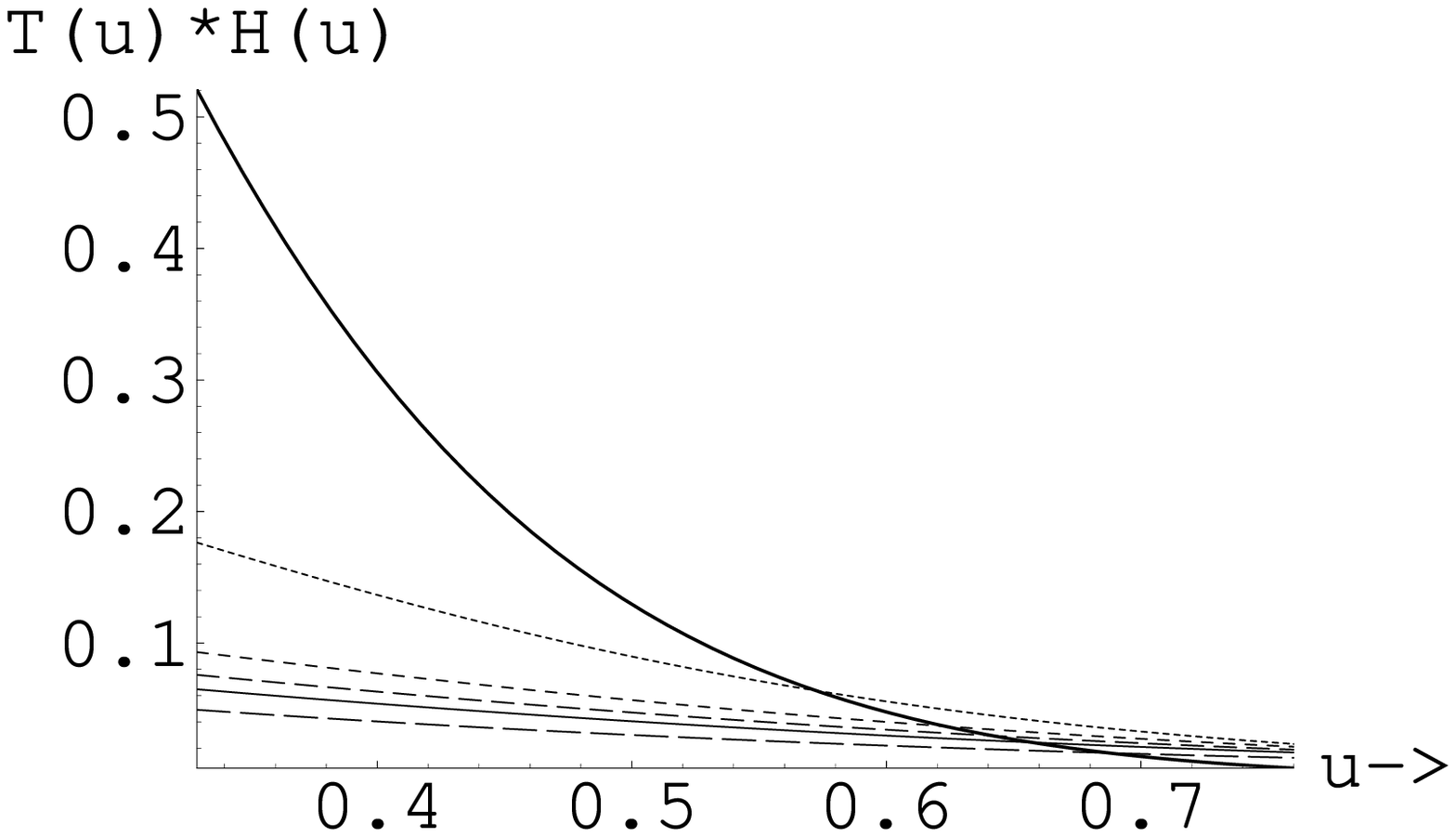}}
\put(70,60){\bf ($\lambda=0.0$, $Q{min}=20.$)}
\put(10,120){\epsfxsize=5cm \epsfbox{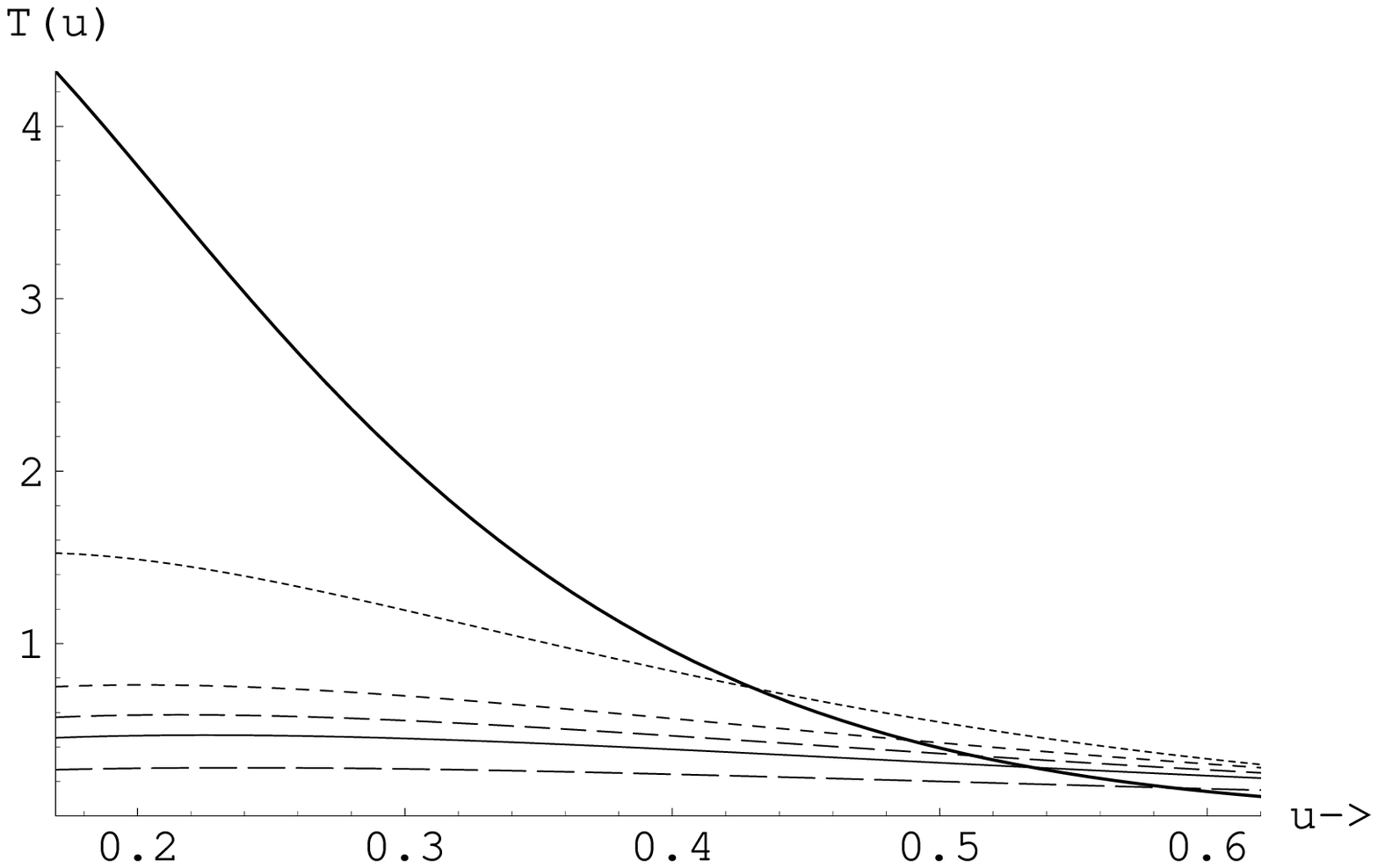}}
\put(10,120){\bf ($\lambda=0.0$, $Q{min}=10.$)}
\put(70,120){\epsfxsize=5cm \epsfbox{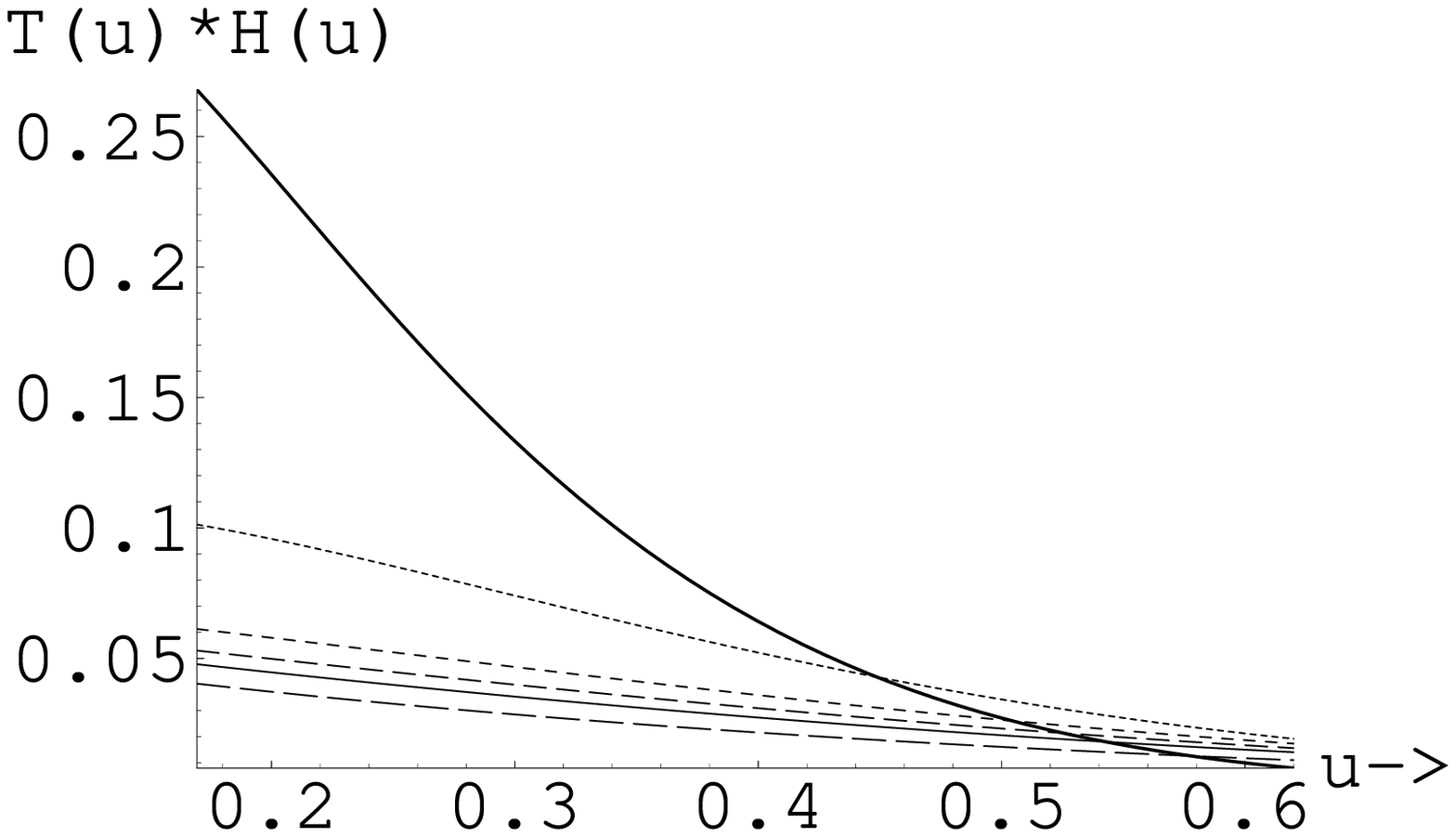}}
\put(70,120){\bf ($\lambda=0.0$, $Q{min}=10.$)}
\put(10,180){\epsfxsize=5cm \epsfbox{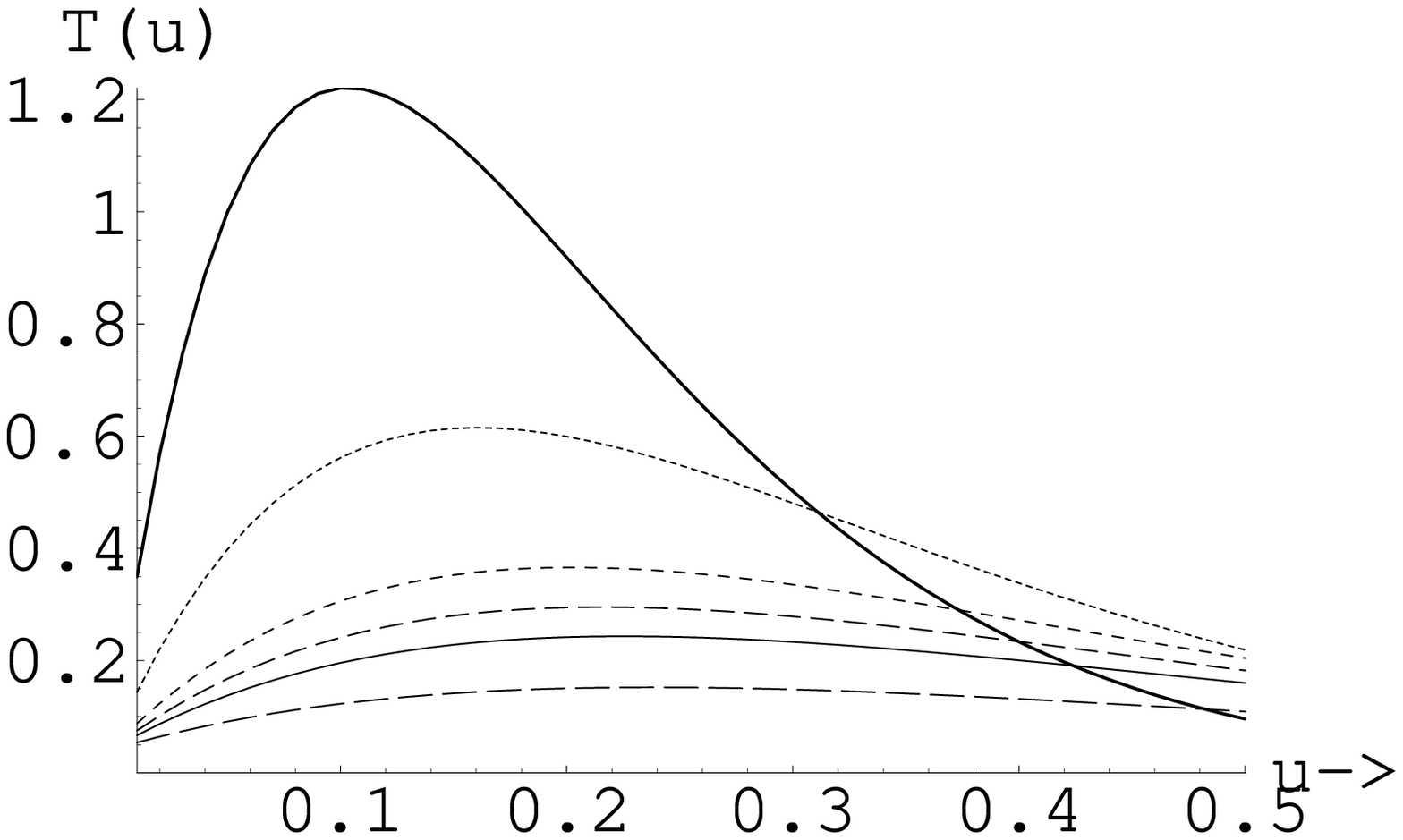}}
\put(10,180){\bf ($\lambda=0.0$, $Q{min}=00.$)}
\put(70,180){\epsfxsize=5cm \epsfbox{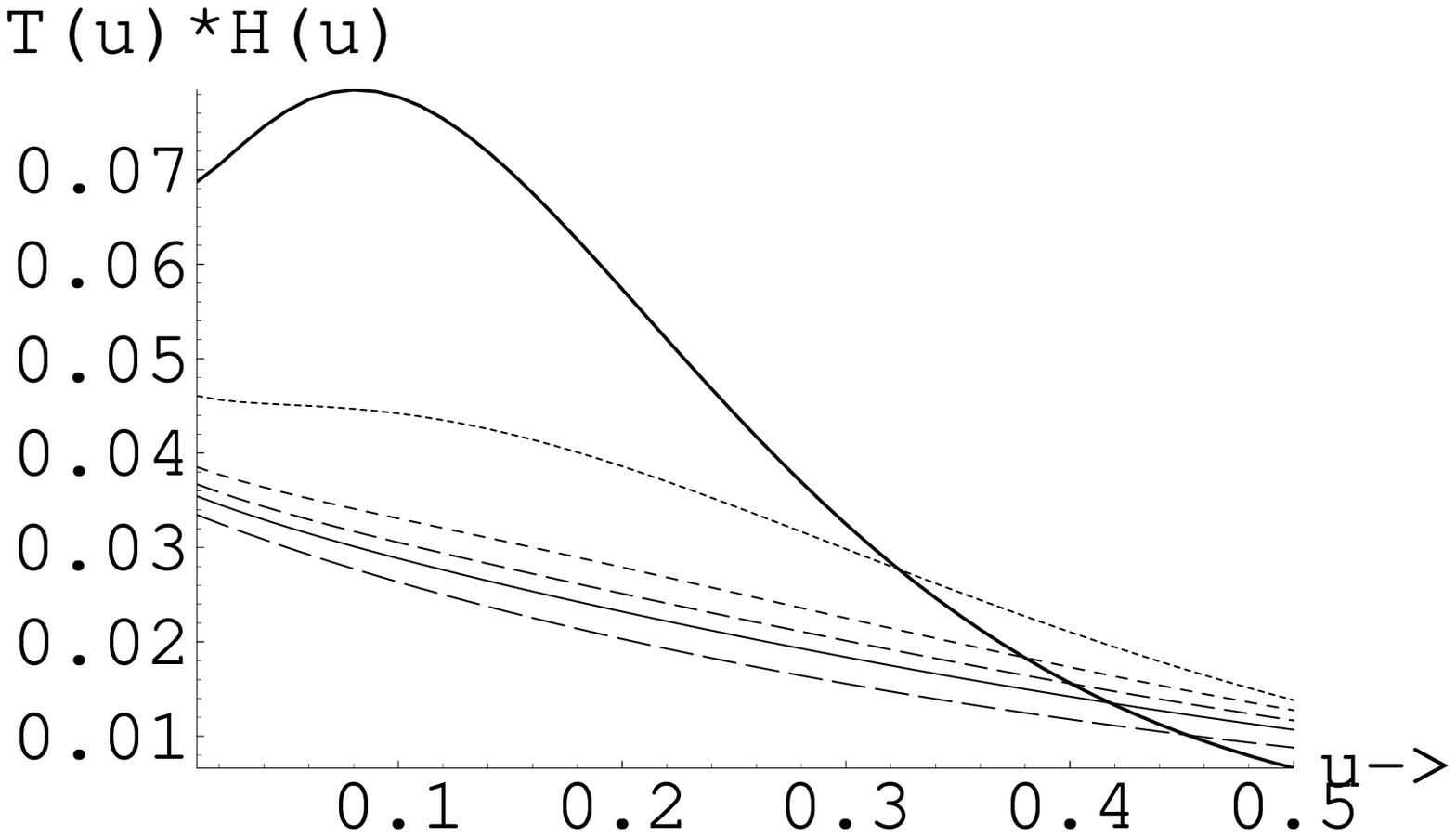}}
\put(80,180){\bf ($\lambda=0.0$, $Q{min}=00.$)}
\end{picture}
\caption[]{The quantities $T,H$ and $TH$ entering the undirectional 
differential rate for $\lambda=0.0$ and various values of energy cit off in 
$GeV$ . For definitions see text. The energy trasfer $Q$ is given by $Q=uQ_0$, 
$Q_0=60KeV$}
\label{fig.1}
\end{figure}
\setlength{\unitlength}{1mm}
\begin{figure}
\begin{picture}(150,200)
\put(40,0){\epsfxsize=5cm \epsfbox{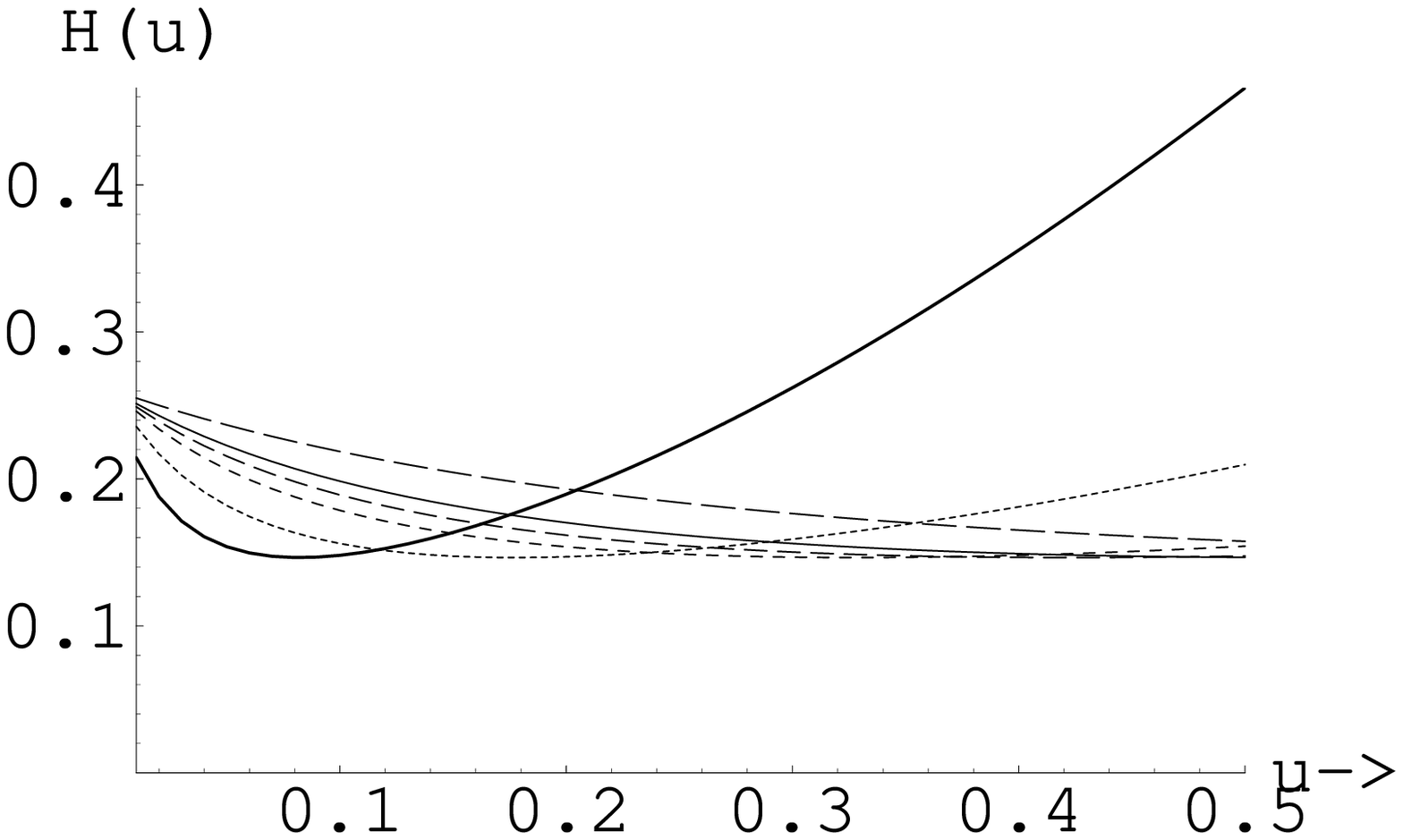}}
\put(40,0){\bf ($\lambda=0.5$, independent of $Q_{min}$)}
\put(10,60){\epsfxsize=5cm \epsfbox{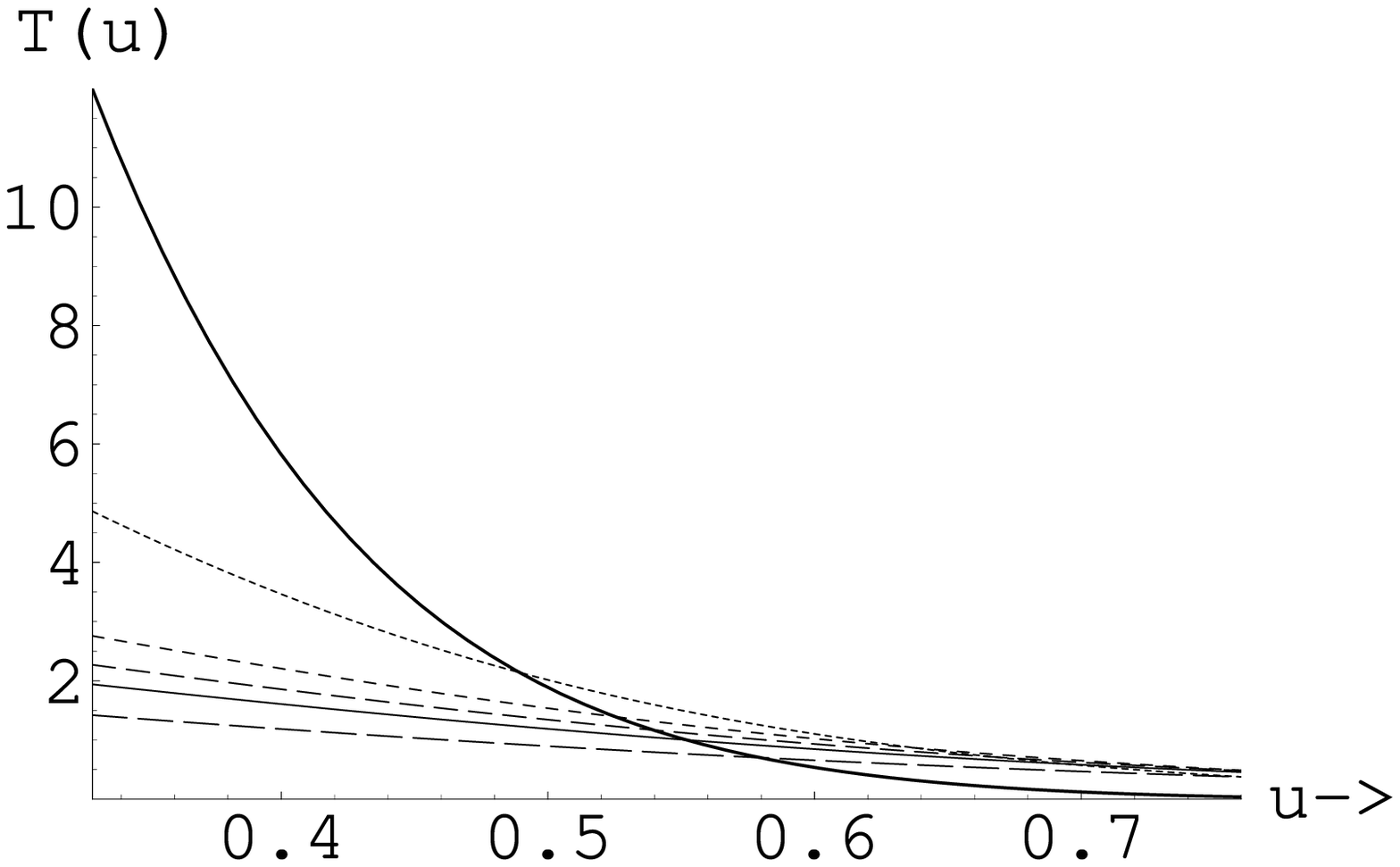}}
\put(10,60){\bf ($\lambda=0.5$, $Q_{min}=20.0$)}
\put(70,60){\epsfxsize=5cm \epsfbox{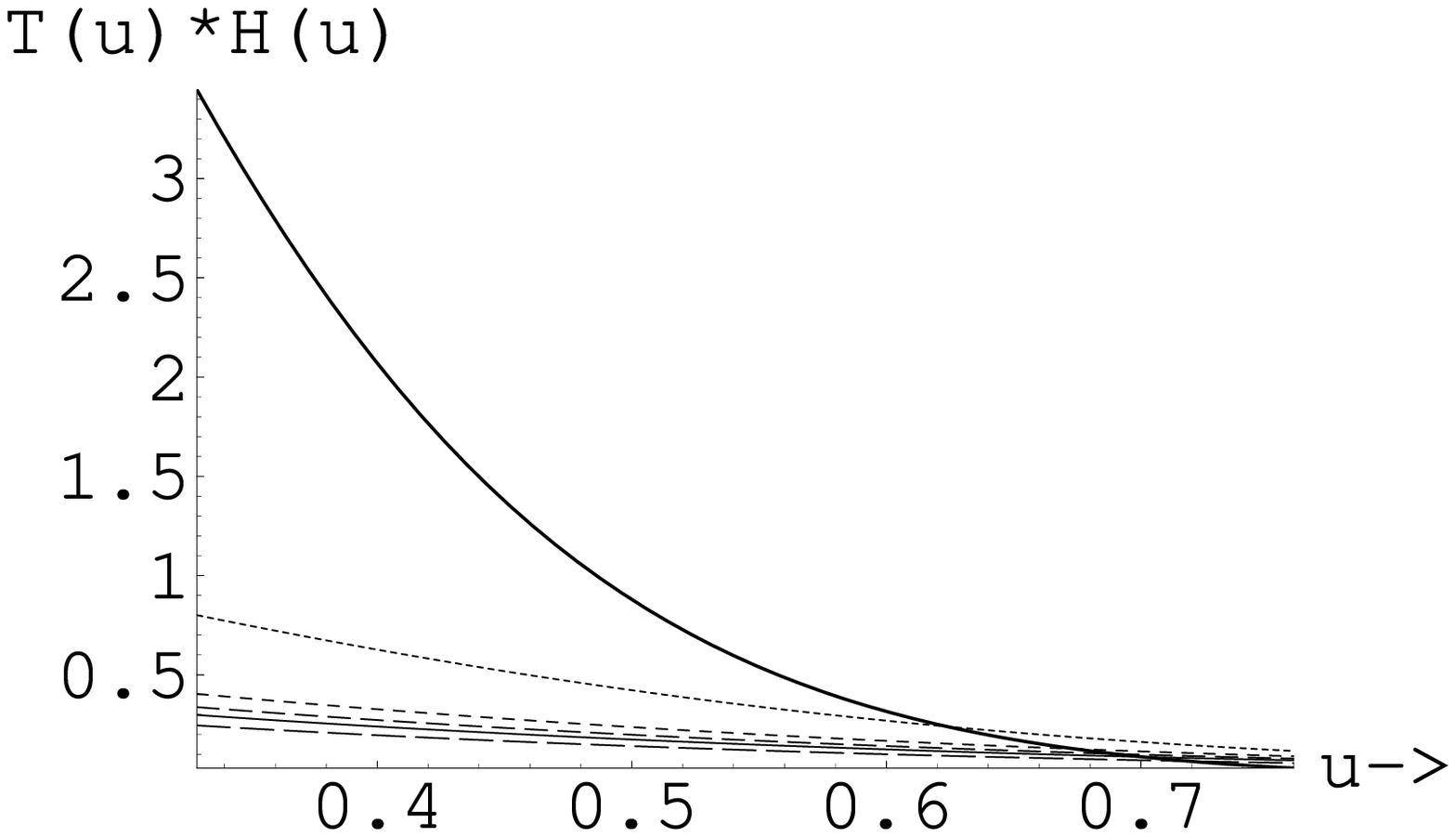}}
\put(70,60){\bf ($\lambda=0.5$, $Q{min}=20.$)}
\put(10,120){\epsfxsize=5cm \epsfbox{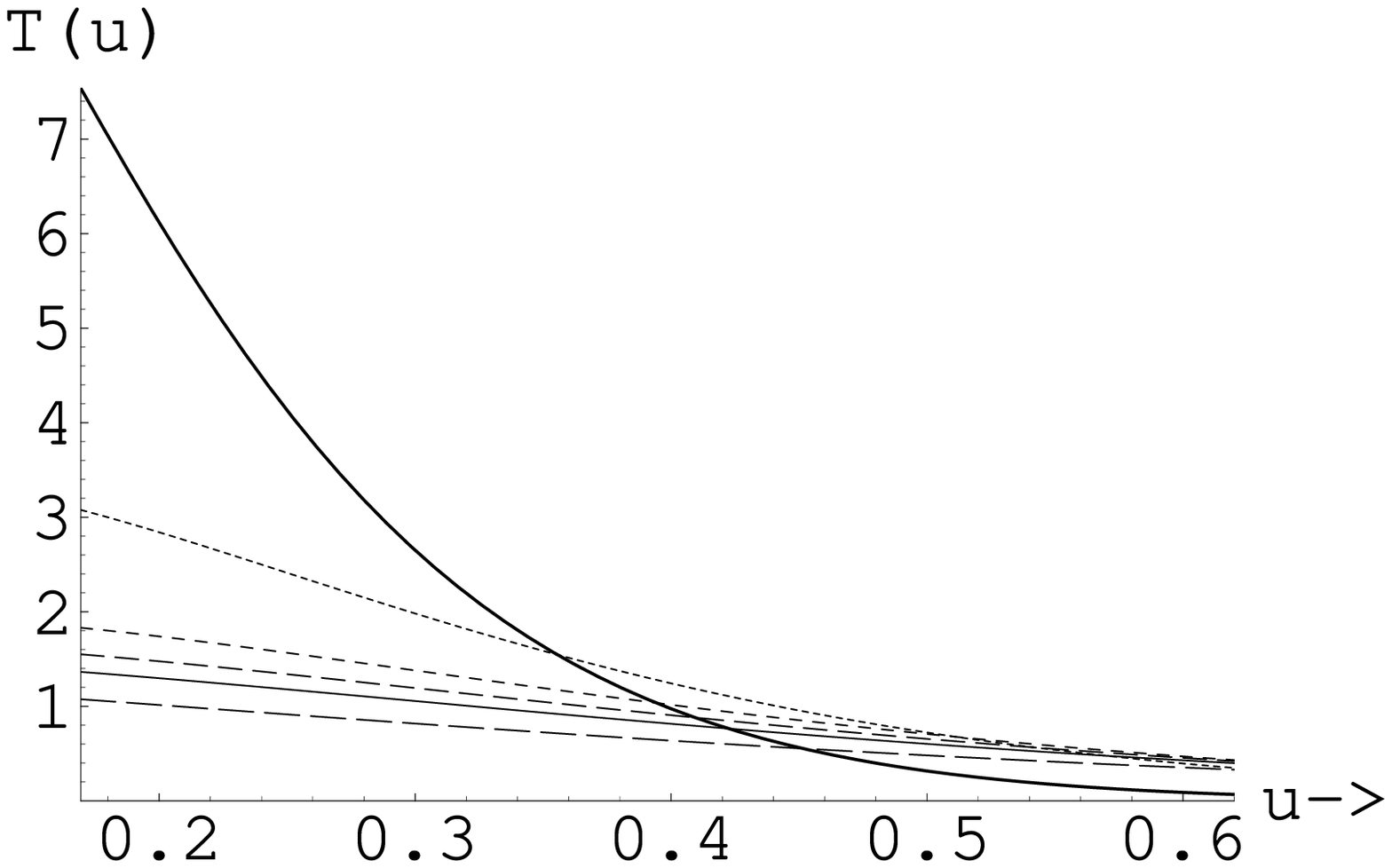}}
\put(10,120){\bf ($\lambda=0.5$, $Q{min}=10.$)}
\put(70,120){\epsfxsize=5cm \epsfbox{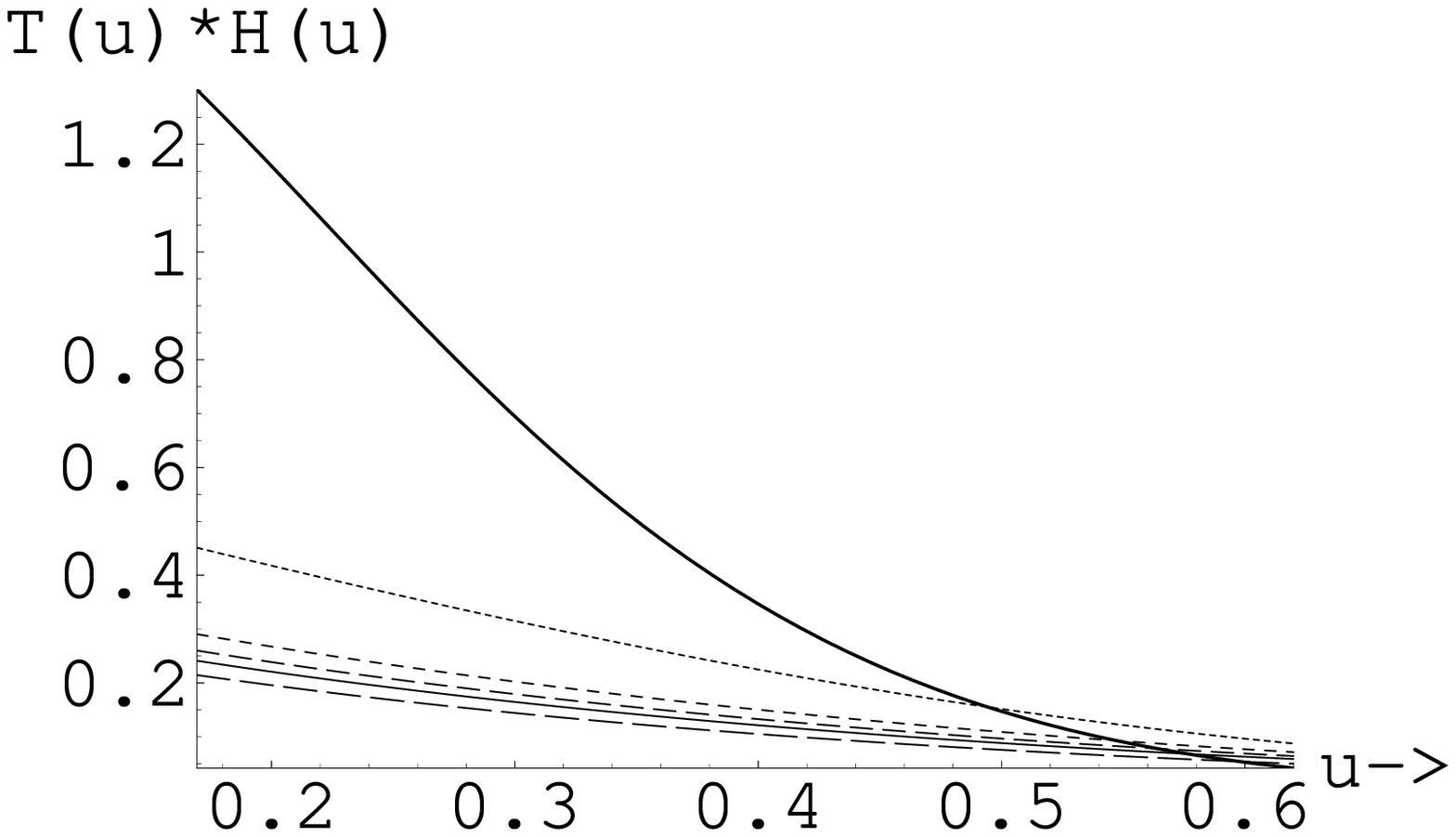}}
\put(70,120){\bf ($\lambda=0.5$, $Q{min}=10.$)}
\put(10,180){\epsfxsize=5cm \epsfbox{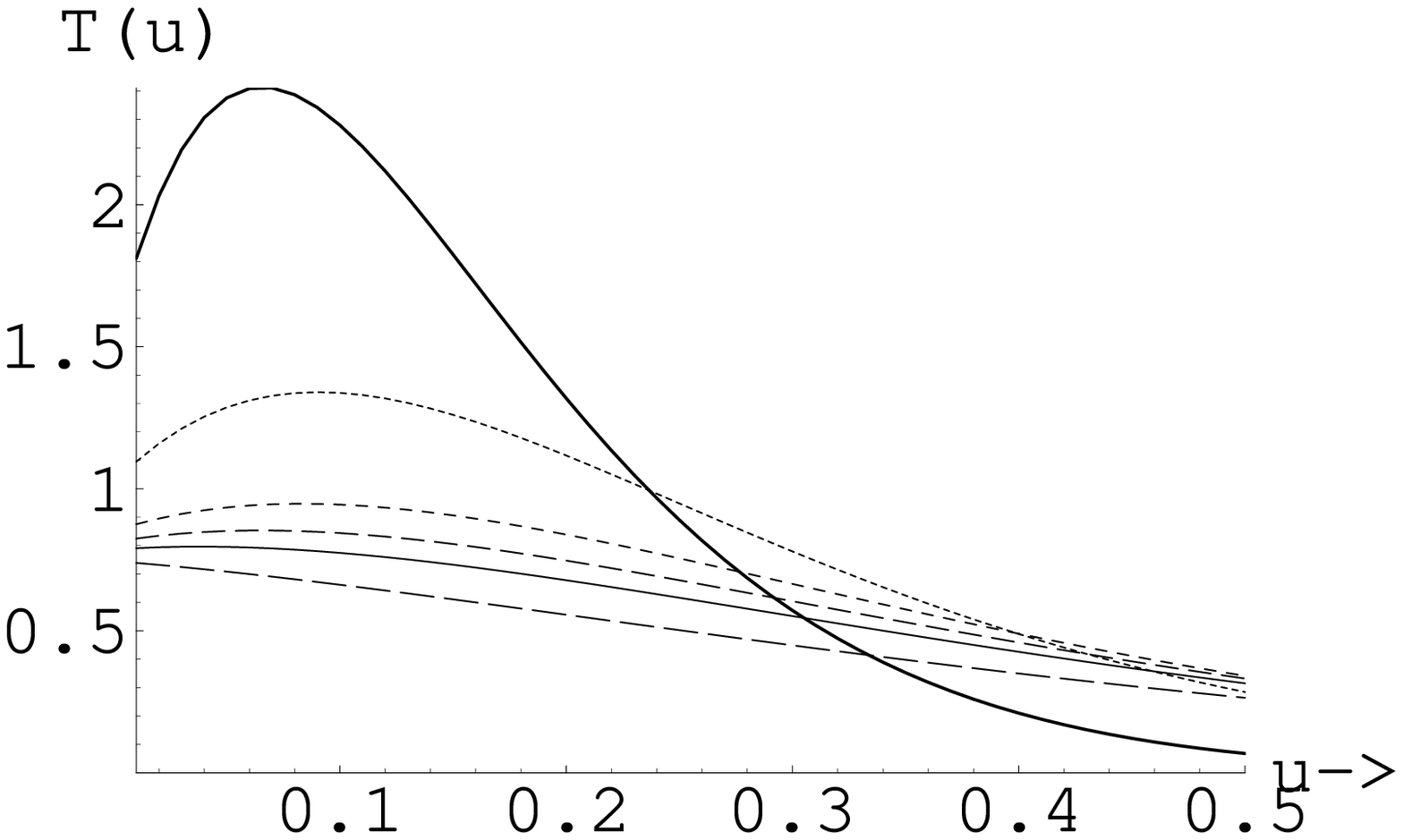}}
\put(10,180){\bf ($\lambda=0.5$, $Q{min}=00.$)}
\put(70,180){\epsfxsize=5cm \epsfbox{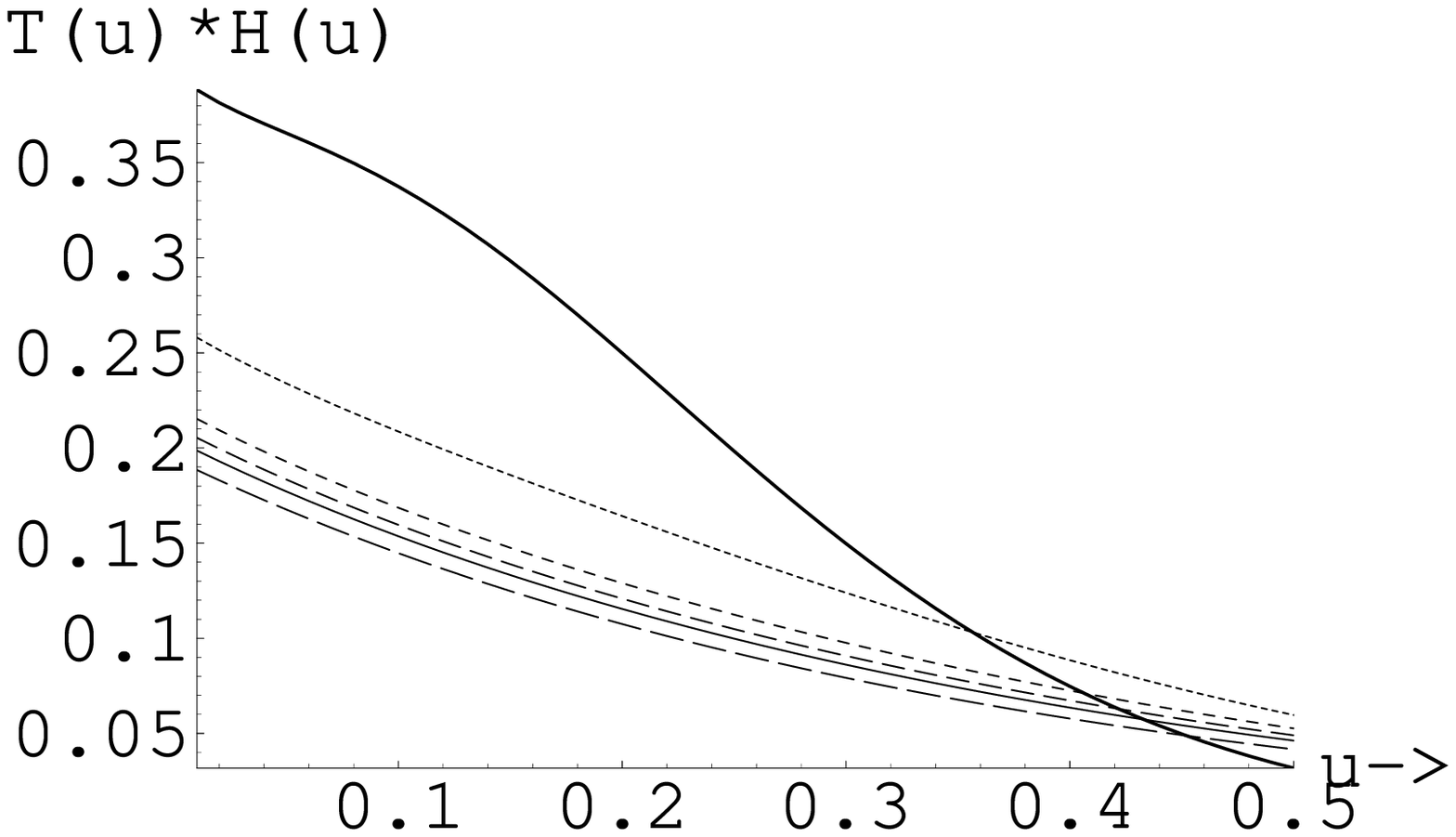}}
\put(70,180){\bf ($\lambda=0.5$, $Q{min}=00.$)}
\end{picture}
\caption[]{The same as in the previous figure for $\lambda=0.5$}
\label{fig.2}
\end{figure}
\setlength{\unitlength}{1mm}
\begin{figure}
\begin{picture}(150,200)
\put(40,0){\epsfxsize=5cm \epsfbox{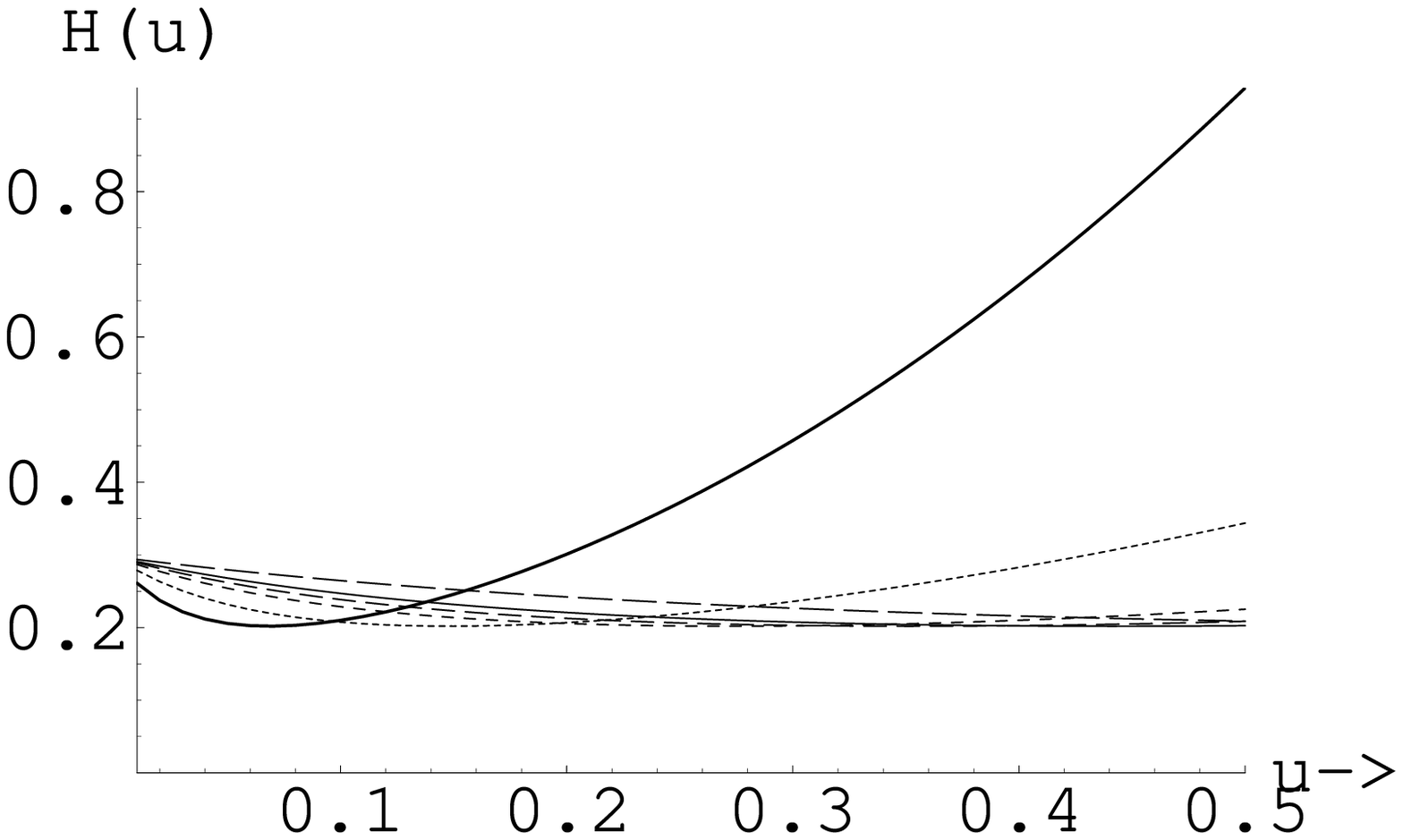}}
\put(40,0){\bf ($\lambda=1.0$, independent of $Q_{min}$)}
\put(10,60){\epsfxsize=5cm \epsfbox{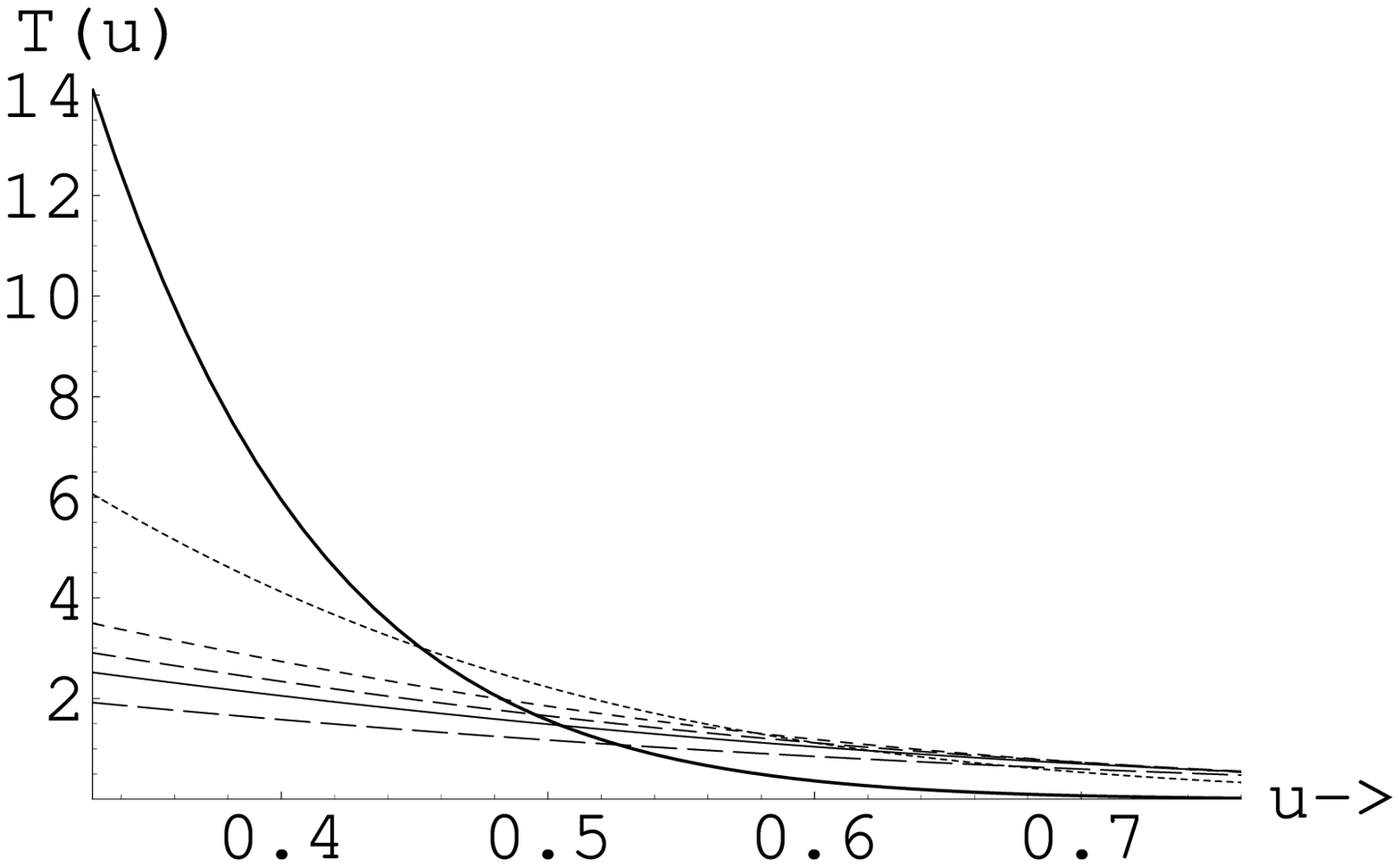}}
\put(10,60){\bf ($\lambda=1.0$, $Q_{min}=20.0$)}
\put(70,60){\epsfxsize=5cm \epsfbox{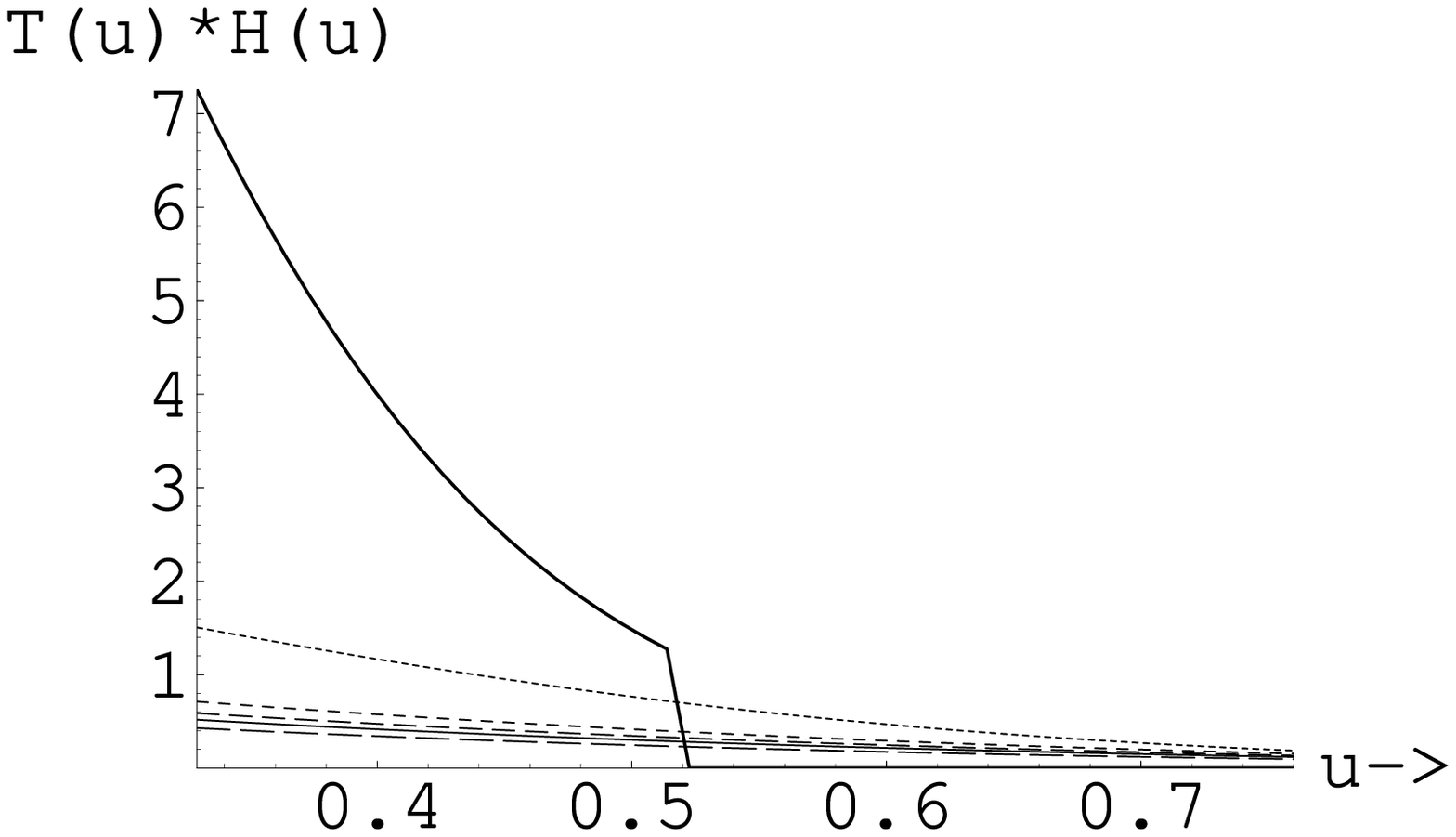}}
\put(70,60){\bf ($\lambda=1.0$, $Q{min}=20.$)}
\put(10,120){\epsfxsize=5cm \epsfbox{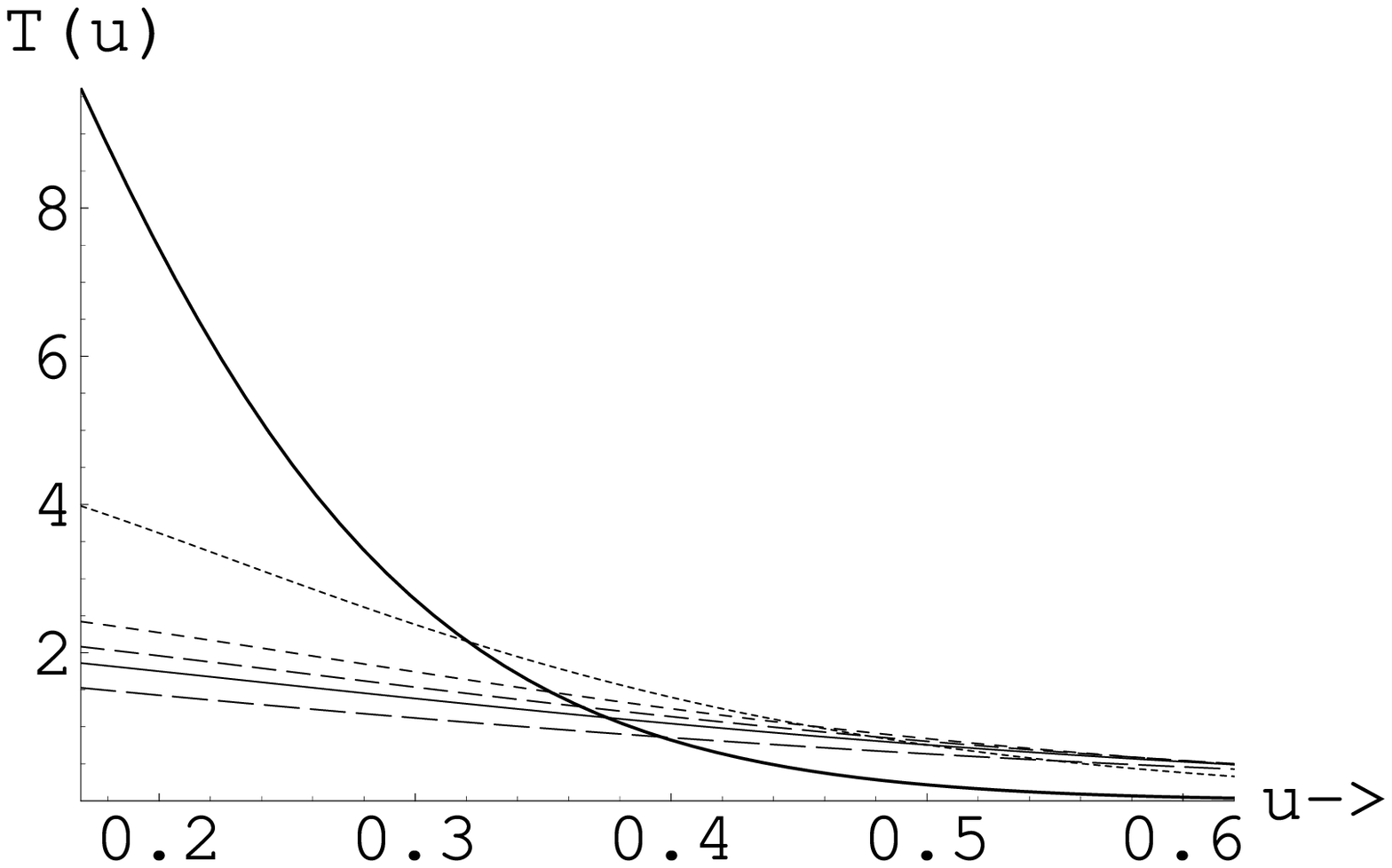}}
\put(10,120){\bf ($\lambda=1.0$, $Q{min}=10.$)}
\put(70,120){\epsfxsize=5cm \epsfbox{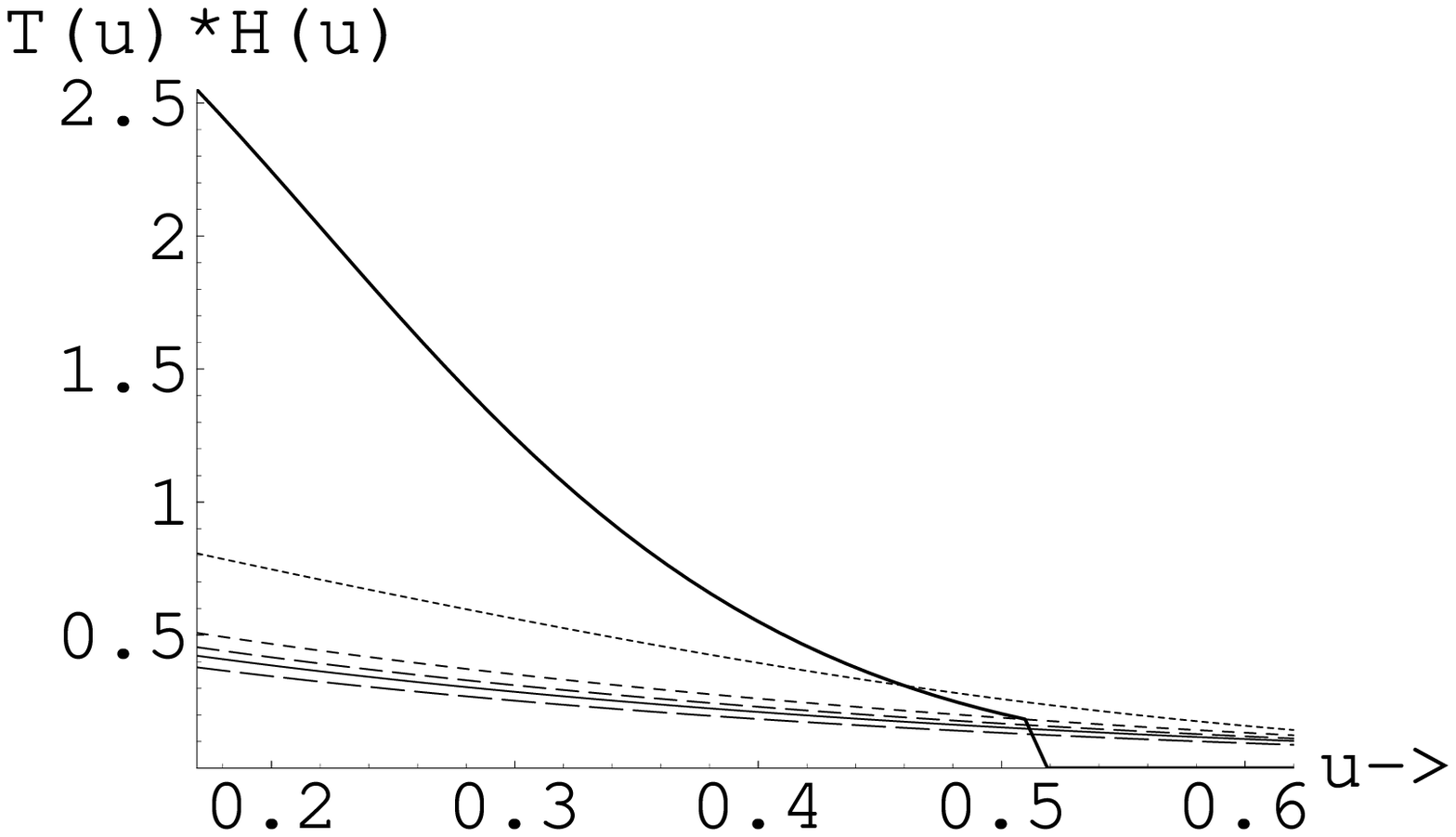}}
\put(70,120){\bf ($\lambda=1.0$, $Q{min}=10.$)}
\put(10,180){\epsfxsize=5cm \epsfbox{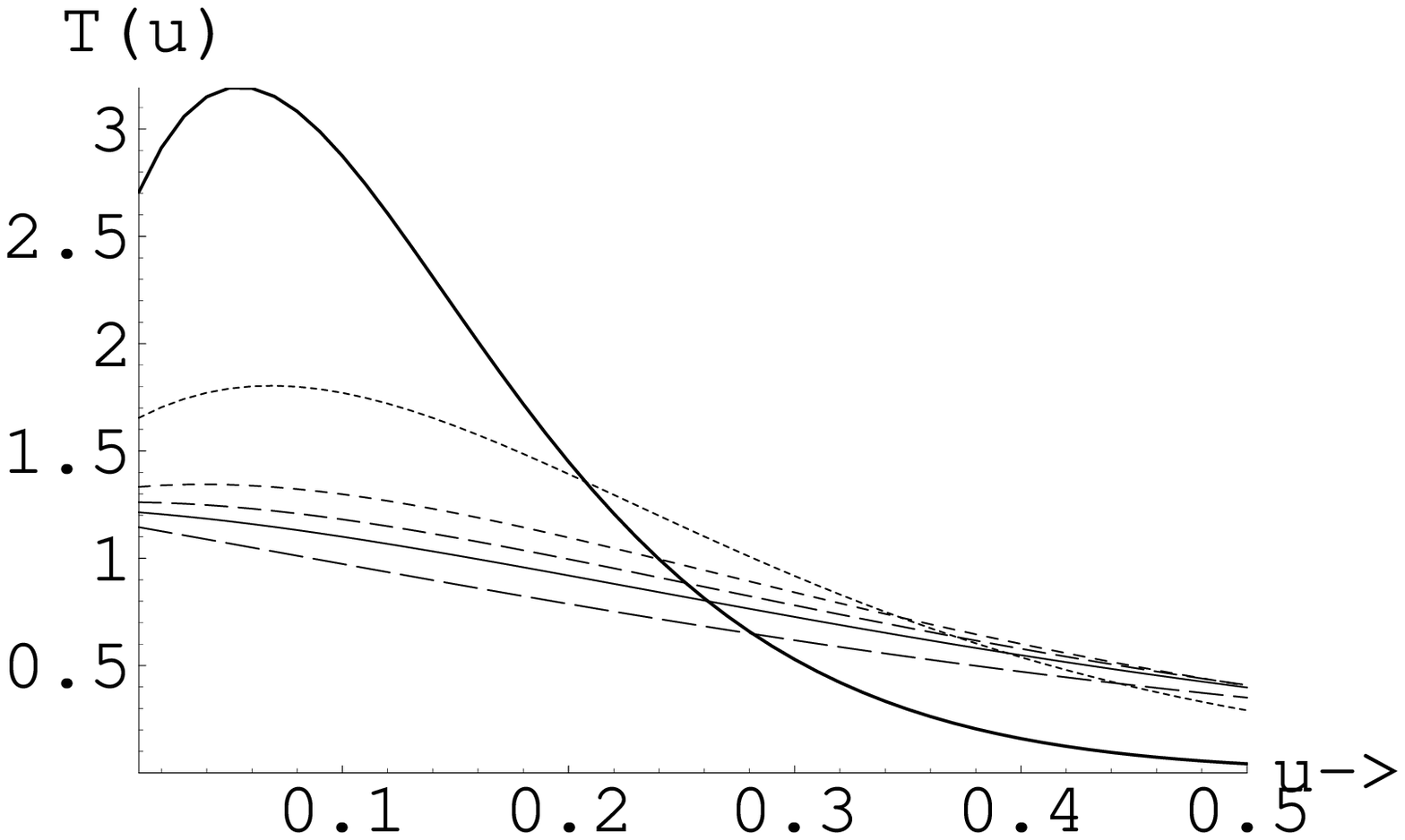}}
\put(10,180){\bf ($\lambda=1.0$, $Q{min}=00.$)}
\put(70,180){\epsfxsize=5cm \epsfbox{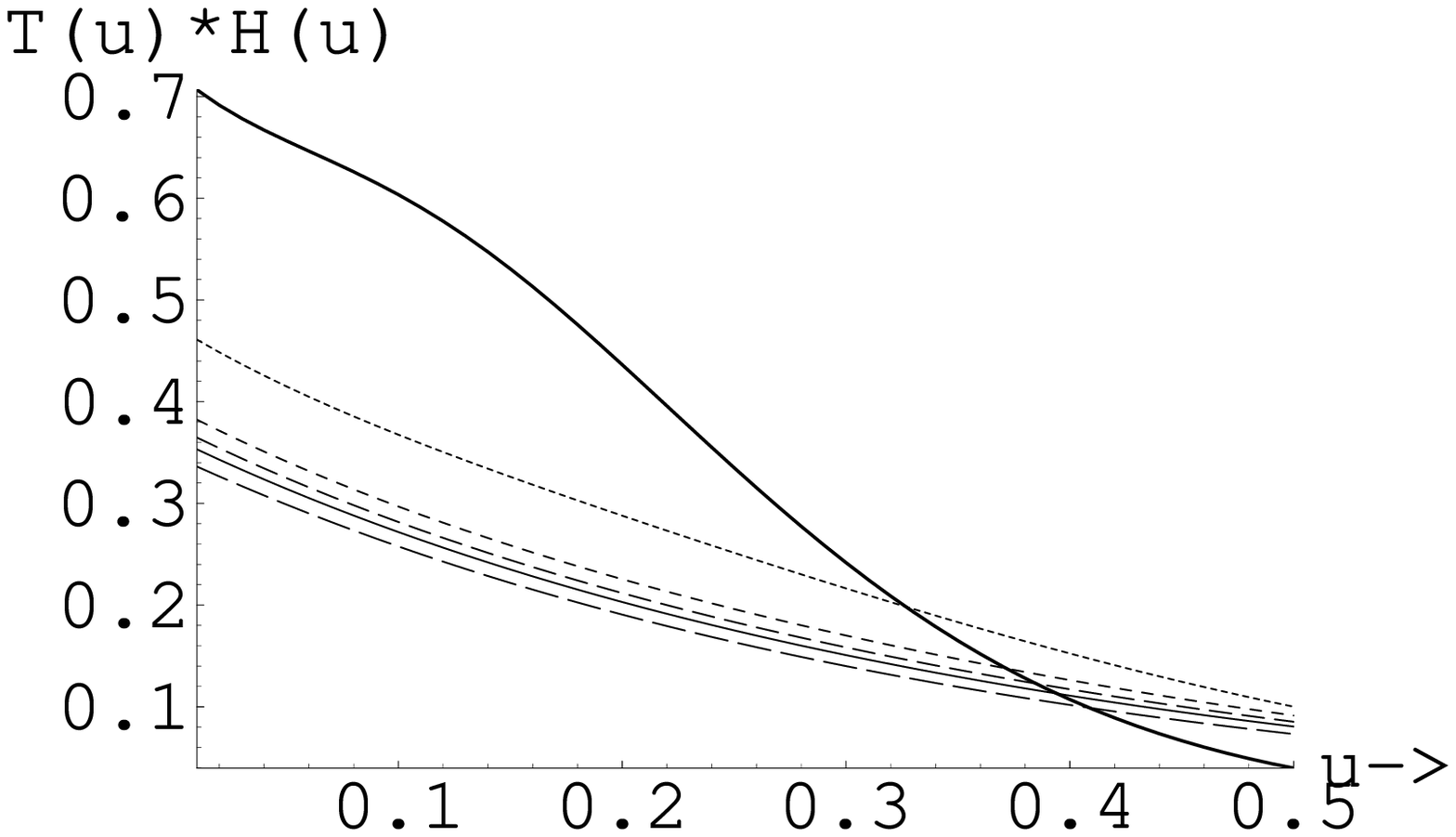}}
\put(70,180){\bf ($\lambda=1.0$, $Q{min}=00.$)}
\end{picture}
\caption[]{The same as in the previous figure for $\lambda=1.0$}
\label{fig.3}
\end{figure}
\\
 In order to understand the dependence of the total and differential rates on
$\lambda$, we will examine the functions
$f(y)$, which are equal to the quantity $N(\lambda,y)~\frac{2}{\sqrt{6\pi}}$ 
multiplied by the integrand of Eq.
(\ref{3.29}). THe latter crucially depends on the functions 
$\tilde{F}_{i}(\lambda,2(\lambda+1)y)$ and $\tilde{G}_{i}(\lambda,y)$ , 
$i=1,2$. 
These functions $f(y)$ are shown in Fig. 4 for $\lambda=0,1$. We see that in 
the case of $\lambda=1.0$ the positive section of the function is enhanced.
\setlength{\unitlength}{1mm}
\begin{figure}
\begin{picture}(150,30)
\put(10,0){\epsfxsize=5cm \epsfbox{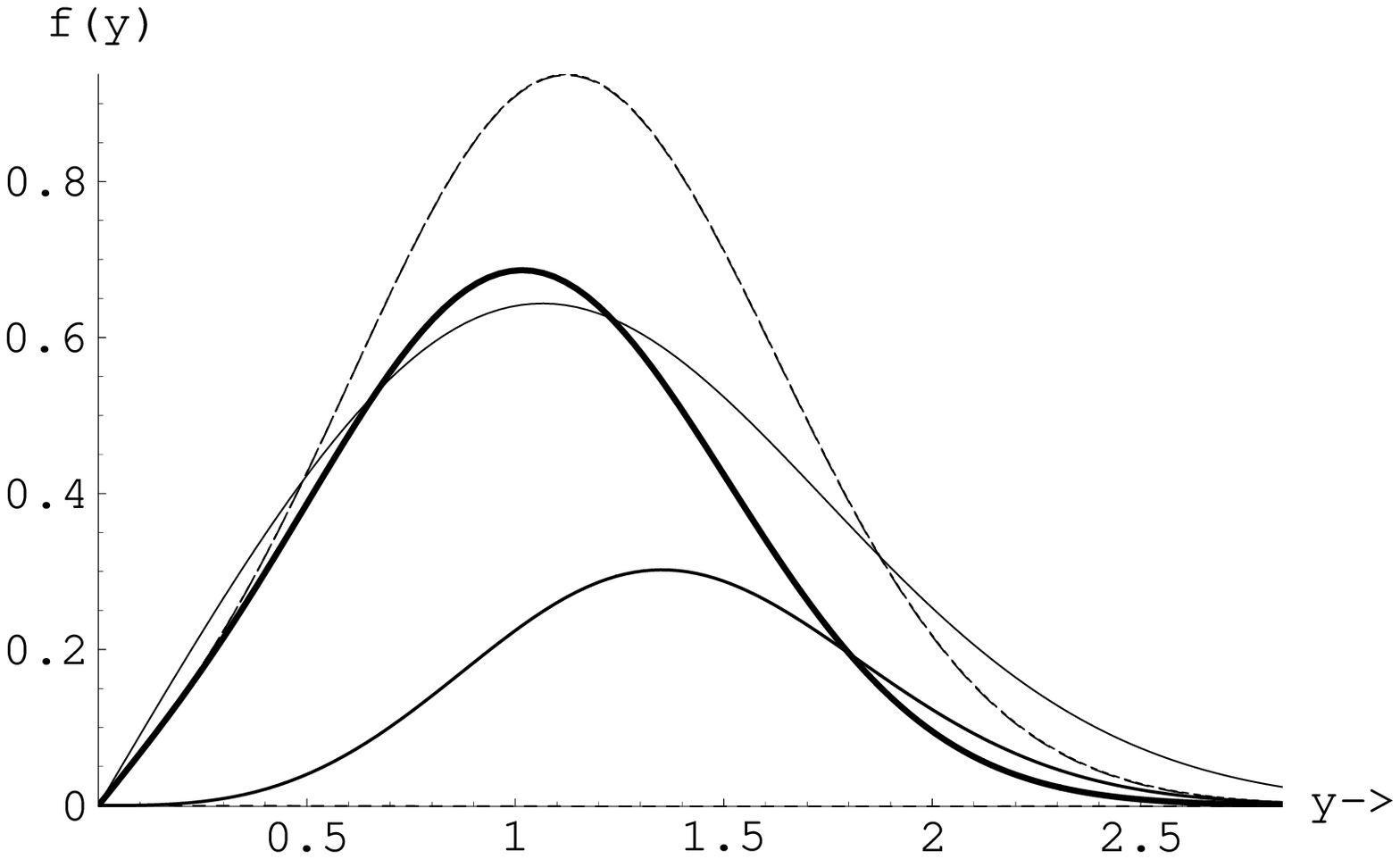}}
\put(10,0){\bf (
 $f(y)\leftrightarrow\tilde{F}_{1},\tilde{G}_{1}$, $\tilde{F}_{1}+\tilde{G}_{1}$
)}
\put(70,0){\epsfxsize=5cm \epsfbox{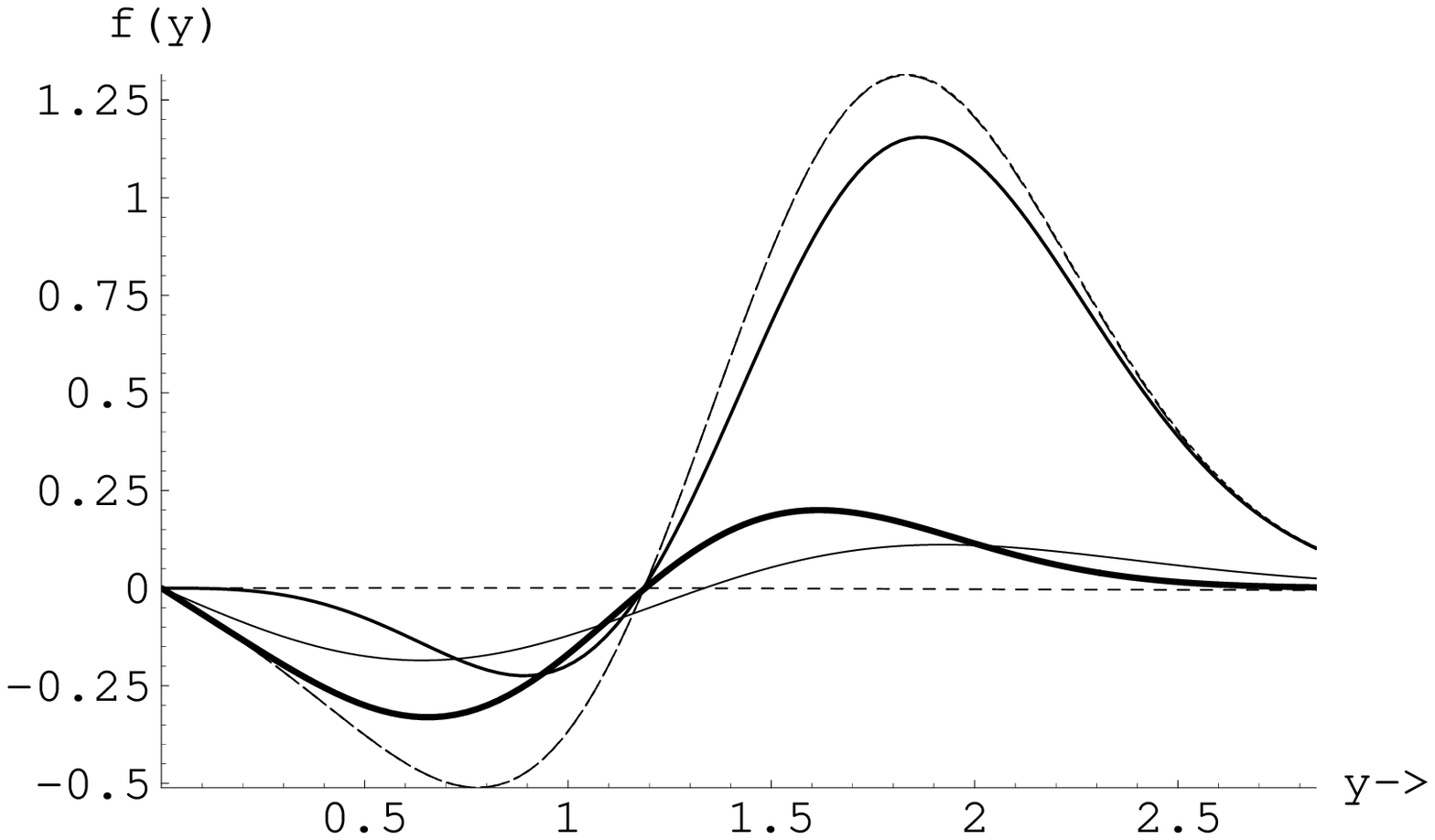}}
\put(70,0){\bf (
 $f(y)\leftrightarrow\tilde{F}_{2},\tilde{G}_{2}$, $\tilde{F}_{2}+\tilde{G}_{2}$
)}
\end{picture}
\caption[]{The quantities $f(y)$ described in the text associated with
 $\tilde{F}_{i}(\lambda,2(\lambda+1)y),\tilde{G}_{i}(\lambda,y)$
and $\tilde{F}_{i}(\lambda,2(\lambda+1)y)+\tilde{G}_{i}(\lambda,y)~$,
$~~~i=1,2$.
Thick solid line corresponds to $\tilde{F}_{i}$ $i=1,2$,the finest line to 
$\tilde{G}_{i}$ $i=1,2$ 
and the dashed line to the sum of the two. The intemediate thickness line
corresponds to $\lambda=0$, in which case $\tilde{G}_{i}=0$, $i=1,2$}
\label{fig.4}
\end{figure}
\subsection{The Directional Rates}

 Once again we distinguish two cases, the total and the differential rates.\\

6.2.1 {\it The Directional Total Event Rates}\\

 The directional total event rates, which arize by summing the directional
 rates in all three directions is beyond the goals of the present experiments.
We will, however, include it in the present discussion. The unmodulated rates 
can be can be parameterized in terms of the parameter $t^0$.
This describes the modification of the total directional
non modulated event rate due to the convolution with the velocity distribution.
The modulation is now described by the three parameters $h_1,h_2,h_3$.  These
are shown in tables 5-7. It is clear from Eq. (\ref{4.56}) that the modulation
of the total rate is no longer given by a simple  sinusoidal function. 
For some interesting cases the situation is shown in Fig. 5.
\begin{figure}
\vspace*{-0.5cm}
\centerline
{\vbox{\psfig{figure=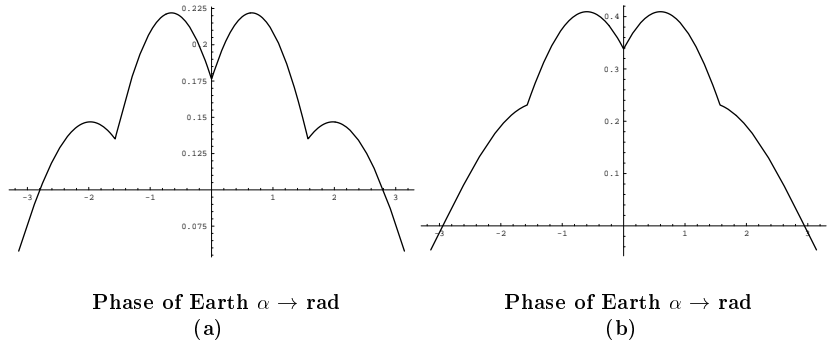,height=20cm
}}}
\vspace*{-14.0cm}
\caption[]{
The the total directional modulation amplitude as a function of the phase of 
the earth
$\alpha$ in two cases.a) for $h_1=0.059$, $h_2=0.117$ and $h_3=0.135$ and
b) $h_1=0.192$, $h_2=0.146$ and $h_3=0.231$. Note that in case b the minimum 
is negative. The results shown are for the target   
$_{53}I^{127}$ (for the definitions see text). 
}
\label{fig.5}
\end{figure}

An idea about what is happening can be given by $h_m$ and $\alpha_m$. The first
gives the difference between the maximum and the minimum values of modulated
amplitude. The second involves the phase shift from the second of June, which 
is no longer the the date of the maximum. 



The second of June gives the location
of the maximum, when only the component of the earth`s velocity along
the sun`s direction of motion is considered or when $h_3$ is neglected. In 
almost all cases considered in this work, however, $h_3$ is important and in
fact  the obtained shift is on the average about $\pm$ 35 days from the second
of June.

 From  tables 5-7 we see that without detector energy cutoff, $Q_{min}=0$, $t^0$
is a decreasing value of 
the LSP mass. It takes values of about 2 for low LSP mass and is decreased by an
order of magnitude as we go to higher LSP masses. This is due to the nuclear
form factor effects, which are present but not so severe for this intermediate
mass nucleus.
 It is not a sensitive function of the asymmetry parameter $\lambda$.

\begin{table}[t]  
\caption{The quantities $t^{0},h_1$ and $h_m$ for $\lambda=0$ in the case of the
target $_{53}I^{127}$ for various LSP masses and $Q_{min}$ in KeV (for
definitions see text). Only the scalar contribution is considered. Note that in
this case $h_2$ and $h_3$ are constants equal to 0.117 and 0.135 respectively.}
\begin{center}
\footnotesize
\begin{tabular}{|l|c|rrrrrrr|}
\hline
\hline
& & & & & & & &     \\
&  & \multicolumn{7}{|c|}{LSP \hspace {.2cm} mass \hspace {.2cm} in GeV}  \\ 
\hline 
& & & & & & & &     \\
Quantity &  $Q_{min}$  & 10  & 30  & 50  & 80 & 100 & 125 & 250   \\
\hline 
& & & & & & & &     \\
$t^0$ &0.0&1.960&1.355&0.886&0.552&0.442&0.360&0.212 \\
$h_1$ &0.0&0.059&0.048&0.037&0.029&0.027&0.025&0.023 \\
$h_m$ &0.0&0.164&0.144&0.124&0.111&0.107&0.104&0.100 \\
\hline 
& & & & & & & &     \\
$t^0$ &10.&0.000&0.365&0.383&0.280&0.233&0.194&0.119 \\
$h_1$ &10.&0.000&0.086&0.054&0.038&0.033&0.030&0.025 \\
$h_m$ &10.&0.000&0.214&0.155&0.127&0.119&0.113&0.104 \\
\hline 
& & & & & & & &     \\
$t^0$ &20.&0.000&0.080&0.153&0.136&0.11&0.102&0.065 \\
$h_1$ &20.&0.000&0.123&0.073&0.048&0.041&0.036&0.028 \\
$h_m$ &20.&0.000&0.282&0.190&0.145&0.132&0.123&0.109 \\
\hline
\hline
\end{tabular}
\end{center}
\end{table}
\begin{table}[t]  
\caption{The same as in the previous table, but for the value of the asymmetry 
parameter $\lambda=0.5$.}
\begin{center}
\footnotesize
\begin{tabular}{|l|c|rrrrrrr|}
\hline
\hline
& & & & & & & &     \\
&  & \multicolumn{7}{|c|}{LSP \hspace {.2cm} mass \hspace {.2cm} in GeV}  \\ 
\hline 
& & & & & & & &     \\
Quantity &  $Q_{min}$  & 10  & 30  & 50  & 80 & 100 & 125 & 250   \\
\hline 
& & & & & & & &     \\
$t^0$ & 0.0 &2.309&1.682&1.153&0.737&0.595&0.485&0.288 \\
$h_1$ & 0.0 &0.138&0.128&0.117&0.108&0.105&0.103&0.100\\
$h_2$ & 0.0 &0.139&0.137&0.135&0.133&0.133&0.133&0.132\\
$h_3$ & 0.0 &0.175&0.171&0.167&0.165&0.163&0.162&0.162\\
$h_m$ & 0.0 &0.327&0.307&0.284&0.266&0.261&0.257&0.250 \\
\hline 
& & & & & & & &     \\
$t^0$ & 10. &0.000&0.376&0.468&0.365&0.308&0.259&0.160\\
$h_1$ & 10. &0.000&0.174&0.139&0.120&0.114&0.110&0.103\\
$h_2$ & 10. &0.000&0.145&0.138&0.135&0.134&0.134&0.133\\
$h_3$ & 10. &0.000&0.188&0.174&0.167&0.165&0.164&0.162\\
$h_m$ & 10. &0.000&0.400&0.328&0.290&0.278&0.270&0.256 \\
\hline 
& & & & & & & &     \\
$t^0$ & 20. &0.000&0.063&0.170&0.171&0.153&0.134&0.087\\
$h_1$ & 20. &0.000&0.216&0.162&0.133&0.124&0.118&0.107\\
$h_2$ & 20. &0.000&0.155&0.143&0.137&0.136&0.135&0.133\\
$h_3$ & 20. &0.000&0.209&0.182&0.171&0.168&0.166&0.164\\
$h_m$ & 20. &0.000&0.487&0.374&0.316&0.299&0.286&0.265 \\
\hline
\hline
\end{tabular}
\end{center}
\end{table}

\begin{table}[t]  
\caption{The same as in the previous, but for the value of the asymmetry parameter
 $\lambda=1.0$.}
\begin{center}
\footnotesize
\begin{tabular}{|l|c|rrrrrrr|}
\hline
\hline
& & & & & & & &     \\
&  & \multicolumn{7}{|c|}{LSP \hspace {.2cm} mass \hspace {.2cm} in GeV}  \\ 
\hline 
& & & & & & & &     \\
Quantity &  $Q_{min}$  & 10  & 30  & 50  & 80 & 100 & 125 & 250   \\
\hline 
& & & & & & & &     \\
$t^0$ & 0.0 &2.429&1.825&1.290&0.837&0.678&0.554&0.330\\
$h_1$ & 0.0 &0.192&0.182&0.170&0.159&0.156&0.154&0.150 \\
$h_2$ & 0.0 &0.146&0.144&0.141&0.139&0.139&0.138&0.138\\
$h_3$ & 0.0 &0.232&0.222&0.211&0.204&0.202&0.200&0.198\\
$h_m$ & 0.0 &0.456&0.432&0.404&0.382&0.375&0.379&0.361\\
\hline 
& & & & & & & &     \\
$t^0$ & 10. &0.000&0.354&0.502&0.410&0.349&0.295&0.184\\
$h_1$ & 10. &0.000&0.241&0.197&0.174&0.167&0.162&0.154\\
$h_2$ & 10. &0.000&0.157&0.146&0.142&0.140&0.140&0.138\\
$h_3$ & 10. &0.000&0.273&0.231&0.213&0.208&0.205&0.200\\
$h_m$ & 10. &0.000&0.565&0.464&0.413&0.398&0.387&0.370\\
\hline 
& & & & & & & &     \\
$t^0$ & 20. &0.000&0.047&0.169&0.186&0.170&0.150&0.100\\
$h_1$ & 20. &0.000&0.297&0.226&0.190&0.179&0.172&0.159\\
$h_2$ & 20. &0.000&0.177&0.153&0.144&0.142&0.141&0.139\\
$h_3$ & 20. &0.000&0.349&0.256&0.224&0.216&0.211&0.203\\
$h_m$ & 20. &0.000&0.709&0.550&0.448&0.424&0.408&0.380\\
\hline
\hline
\end{tabular}
\end{center}
\end{table}

 As expected, in the presence of energy cut off, $t^0$ is greatly reduced and
becomes unobservable for light LSP.  As the LSP mass increases $t^0$ increases.
(see tables 5-7. It reaches a maximum
at about 80 GeV and it starts decreasing. But even for the heaviest LSP this
reduction is not much larger than 1/3, even for $Q_{min}=20KeV$. In other words
a heavy LSP can cause sufficient energy transfer
to partially compensate for the loss of phase space.

We also see that in all cases $h_m$ is much larger than $2 h_1$, suggesting that
when it comes to directional detection all components of the earth's velocity
are important. We also notice that for those LSP masses, which give a
detectable total rate, the modulation amplitude does not appreciably change with
the LSP mass. It is not greatly affected by the energy cutoff $Q_{min}$. It is
enhanced, however, by about a factor of two in going from $\lambda=0$, no
asymmetry, to $\lambda=1$, maximum asymmetry.

From the above discussion it is clear that one needs all three modulation
parameters, $h_1,h_2,h_3$.  We remind the 
reader that in Eq (\ref {3.55}) the z-axis has been chosen in the direction of
the sun's velocity, the y-axis is perpendicular to the plane of the galaxy and
the x-axis in the radial (galactocentric) direction. We mention again that 
$h_2$ and $h_3$ 
are constant, 0.117 and 0.135 respectively, in the symmetric case. On the other
hand $h_1$ and $h_3$ substantially increase in the presence of asymmetry. 

The precise value of the directional rate depends on the direction of
observation. 
One can find optimal orientations, but we are not going to elaborate further.\\

6.2.2 {\it The Directional Differential Event Rates}\\

 The directional differential rate is also very hard to detect, but perhaps
a bit more 
practical than the total rates described in the previous subsection. It can be 
in terms of four functions
of u, namely $R_0(u)$ and $H_i(u), i=1,2,3$. 

 The situation is rather
complicated and following our discussion of the previous section we will give
gross description of the modulation using the functions 
$a_H(u)$ and $H_m(u)$. 
 The phase shift$\alpha _H$ has been found to be a
constant and about 0.7, which corresponds to a shift about  $\pm$35 days
from June the second. 
Since , however, $H_m$ is defined
as the ratio of two quantities, it can appear large because the denominator 
(non modulated rate) becomes small. Thus, following the strategy of the 
previous subsection we also present the quantity
$R_m=R_0 H_m$. These functions  are shown in Fig. 6 for LSP masses in the range
30-250 GeV, $\lambda =0$ and  $Q_{min}=0$,
10KeV and 20KeV. Note that the quantity $H_m$ is itself independent of the
cutoff except that one should look at the u relevant to the allowed  energy
transfer interval. 

\begin{figure}
\vspace*{-0.8cm}
\centerline
{\vbox{\psfig{figure=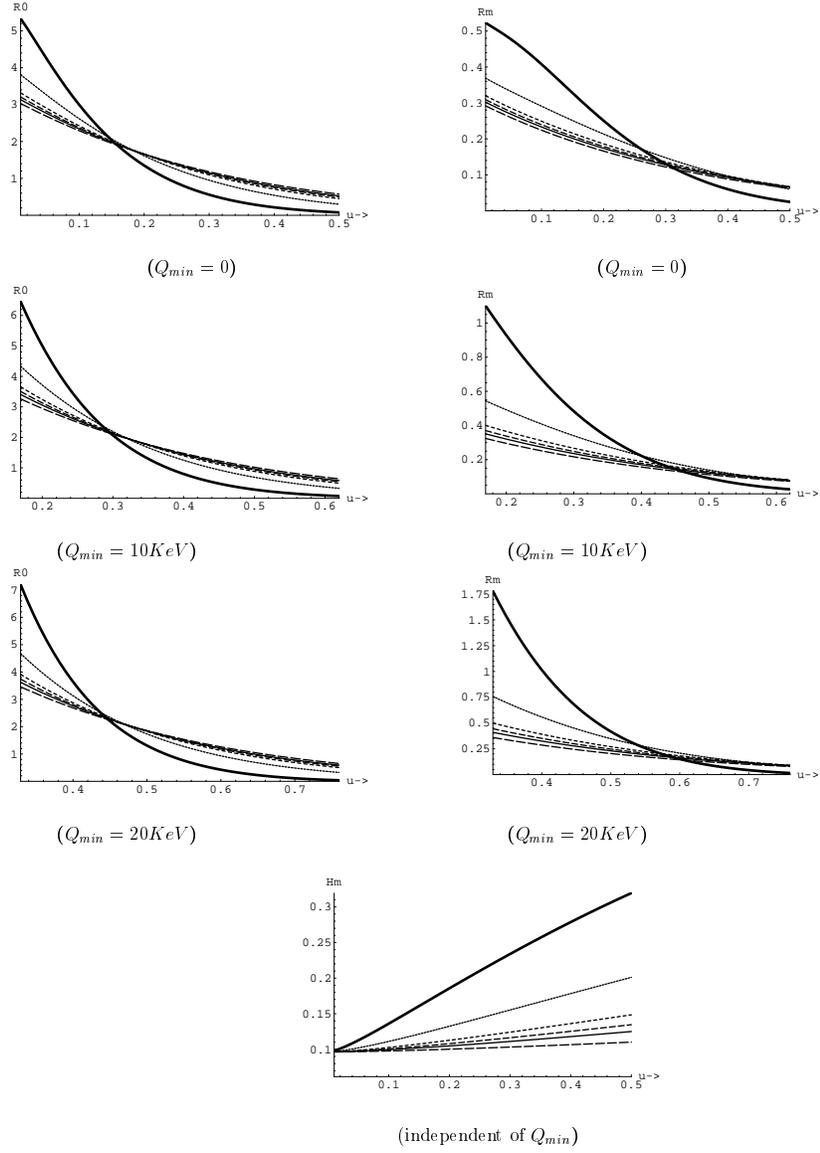,height=7.0in
}}}
\vspace*{-1.3cm}
\caption[]{The relative differential event rate R0 and   
the amplitudes for modulation Rm and Hm vs u for the target 
$_{53}I^{127}$ in the case of symmetric velocity distribution, $\lambda=0$
(for the definitions see text).} 
\label{fig.6}
\end{figure}

The curves shown
correspond  to LSP masses as in the undirectional case.  Again due to the
normalization of R0, the area under the corresponding curve is unity.

 The above quantities for $\lambda=0.5$ and 
1.0 are shown in Fig. 7, but only for $Q_{min}$=0. Their dependence
on the energy transfer cutoff shows behavior similar to that of Fig. 1. 
 In any case for $Q_{min}$=10, 20 KeV, the functions $R_0$ and $R_m$ show a
behavior similar to that for $Q_{min}=0$, except that they start from higher
energy transfer.We should remind the reader, however, that
that in all cases $R_0$ represents the relative differential
rate, i.e. it is normalized so that the area under the corresponding curve
is unity for all LSP masses. One, therefore, should take into account the factor
$t^0$ of tables 5-7. 

\begin{figure}
\vspace*{-3.7cm}
\centerline
{\vbox{\psfig{figure=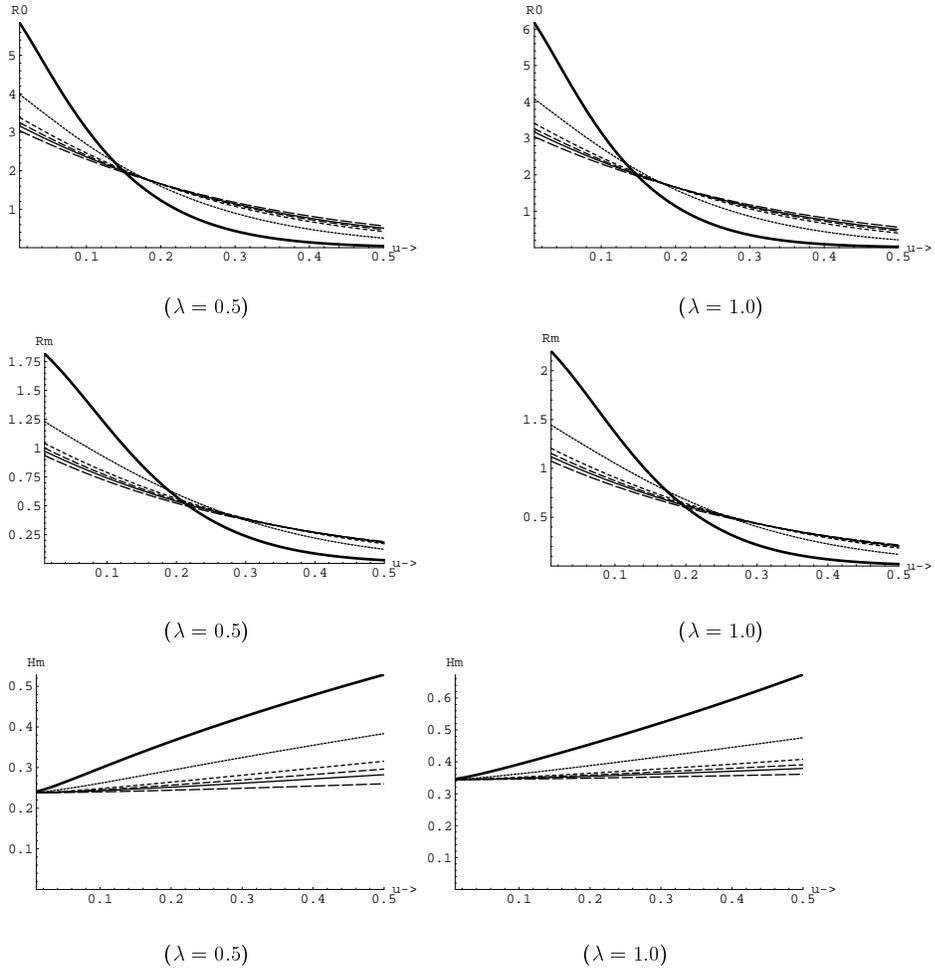,height=8.0in
}}}
\vspace*{-3.5cm}
\caption[]{The same as in Fig.2 in the asymmetric case ($\lambda=0.5$ and
$\lambda=1.0$). Only the case $Q_{min}=0$ is exhibited.}   
\label{fig.8}
\end{figure}

From the plots shown one can see that 
 $R_m$ can be quite large, about 20 $\%$ of $R_0$, but it falls slower
as a function of the energy transfer. For this reason the modulation amplitude
$H_m$ is increasing as a function of u. It is interesting to note that the
modulation amplitude $H_m$ is increased by more than a factor of two, as the
asymmetry
parameter $\lambda$ changes from zero to one, for all energy transfers. Thus,
even at zero energy transfer, for $\lambda=1$ the variation in the amplitude
due to the earth's motion can increase by about $40\%$ between the minimum
(around December 2) and the maximum (around July 10 or the end of may), a big
effect indeed.
It can become even larger if one can restrict oneself to only part of the
phase space, i.e. if one is satisfied with fewer counts.

\begin{figure}
\vspace*{-0.5cm}
\centerline
{\vbox{\psfig{figure=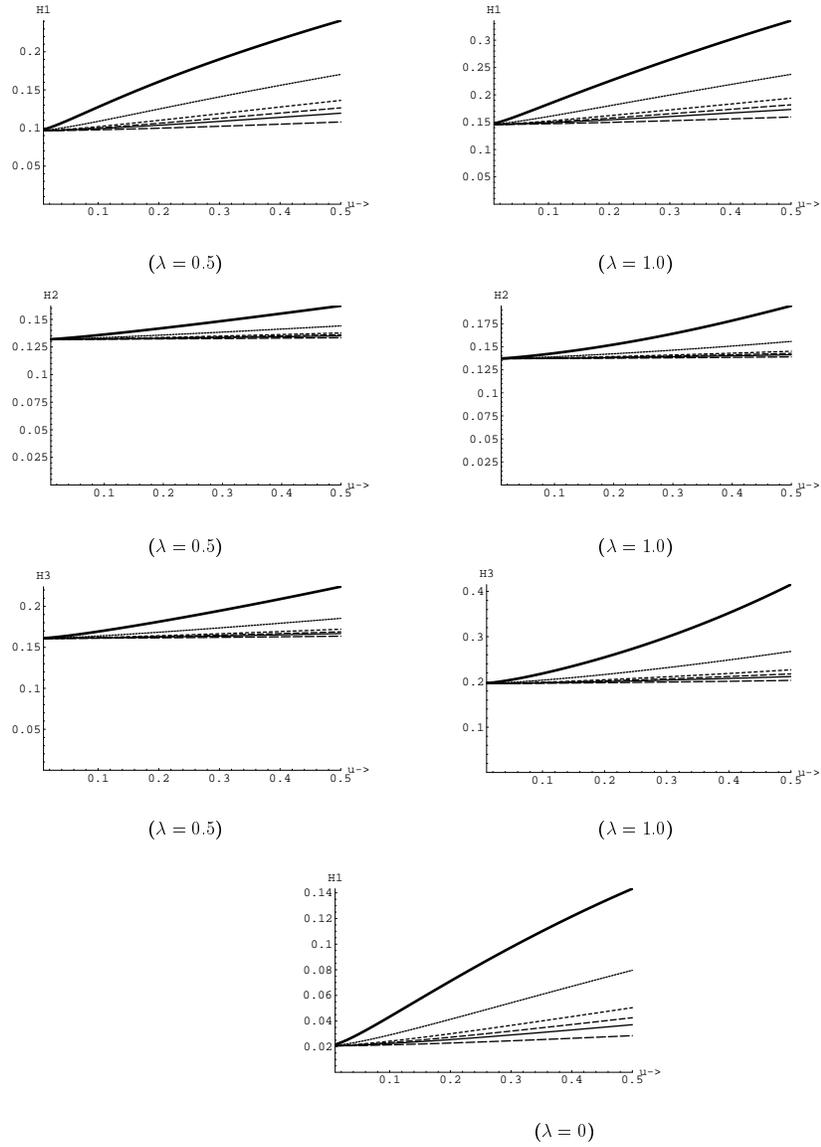,height=7.0in
}}}
\vspace*{-1.5cm}
\caption[]{The same as in Fig.3 for the quantities $H_1,H_2$ and $H_3$. These
quantities do not depend on $Q_{min}$,except for the fact one should look
at $u>u_{min}$.}   
\label{fig.9}
\end{figure}

In the case of the directional differential rate one clearly needs, in addition
to $R_0$ , the functions $H_l(u)$,
l=1,2,3, which are plotted in Fig.4. In the case of $\lambda=0$ only $H_1$ 
is plotted, since the other two are in this case
constant ($H_2=0.117$, $H_3=0.135$). We see that in the presence of asymmetry,
 e.g. $\lambda=0.5$ and 1.0, all functions, but especially $H_1$  
and $H_3$, are substantially increased.


\section{Conclusions}
In the present paper we have expanded the the results obtained in of our recent
letter \cite{JDV99}. We have calculated all the parameters, which can describe 
the modulation of the direct detection rate for supersymmetric dark matter.
 The differential as well as the total event rates were obtained
both for the non directional as well as directional experiments. All components
of the earth's velocity were taken into account, not just its component along
the sun's direction of motion. Realistic axially symmetric velocity
distributions, with enhanced dispersion in the galactocentric direction,
were considered. The obtained results were compared to the up to now
employed Maxwell-Boltzmann distribution.

We presented our results in a suitable fashion so that they do not
depend on parameters of supersymmetry other than the the LSP  mass. 
Strickly speaking the obtained results describe the coherent process
in the case of $^{127}I$, but
we do not expect large changes, if the axial current is considered. Recall
that the dependence on supersymmetry is contained in the parameter $\bar{R}$
not discussed in the present paper.
The nuclear form factor was taken into account and the effects of the detector
energy cut off were also considered.

 Our results, in particular the parameters $t$ and $t_0$ (see tables 2-4 and 
5-8) indicate that for large reduced mass, the advantage of $\mu _r$ (see Eqs.
(\ref{2.10})- (\ref{2.12}) is lost when the nuclear form factor and the 
convolution with the velocity distribution are taken into account. 

In the case of the undirectional total event rates we find that in the 
symmetric case
the modulation amplitude for zero energy cutoff is less than 0.07. 
 It gets substantially increased in the case of asymmetric velocity 
distribution 
with largest asymmetry ($\lambda=1$). It can reach values up t0 $.31$. In the 
presence of the 
detector energy cutoff it can increase even further up to $0.46$, but this
occurs at the expense of the total number of counts. The modulation amplitude
in the case of the differential rate is shifted by the asymmetry at higher
energy transfers and, for maximum asymmetry $\lambda=1$, gets about doubled
compared to the symmetric case ($\lambda=0$).
This  amplitude does not depend on the the energy cutoff, but the lower energy
transfers, will, of course, be excluded if such a cutoff exists.

 Analogous conclusions can be drawn about the directional
differential event rate.
The presence of asymmetry more than triples the differential modulation
amplitude (from about $10\%$ to about $35\%$). 
There
exist now regions of the energy transfer such that the modulation amplitude
can become as large as $50\%$. 

 Finally it is important that one should consider all components of
the earth's motion, not just its velocity along the sun's motion, especially
if the directional signals are to be measured.

\section*{Acknowledgments}

The author would like to
acknowledge partial support of this work by $\Pi $ENE$\Delta $ 1895/95 of 
the Greek Secretariat for research, TMR Nos  ERB FMAX-CT96-0090  and  
ERBCHRXCT93-0323 of the European Union. He would like also to thank the Humboldt
Foundation for their award that provided support during the final stages of this
work and Professor Faessler for his hospitality in Tuebingen. 

\end{document}